\documentclass[reqno, eucal]{amsart}

\usepackage[mathscr]{eucal} 
\usepackage[dvips]{graphicx}
\usepackage[dvips]{color}
\usepackage{amsmath,amsfonts,amssymb,amsthm,amscd}
\usepackage{epic}
\usepackage{eepic}
\usepackage{longtable}
\usepackage{array}

\newtheorem{theorem}{Theorem}[section]
\newtheorem{lemma}[theorem]{Lemma}

\newtheorem{proposition}[theorem]{Proposition}

\theoremstyle{definition}

\newtheorem{remark}[theorem]{Remark}

\newcommand{\Z}{{\mathbb Z}}
\newcommand{\R}{{\mathbb R}}
\newcommand{\C}{{\mathbb C}}

\newcommand{\I}{{\mathrm i}}

\newcommand{\ok}{{\rm{\bf k}}}
\newcommand{\OK}{{\rm{\bf K}}}
\newcommand{\am}{{\rm{\bf a}}^{\!-} }
\newcommand{\ap}{{\rm{\bf a}}^{\!+} }
\newcommand{\apm}{{\rm{\bf a}}^{\!\pm} }
\newcommand{\amp}{{\rm{\bf a}}^{\!\mp} }
\newcommand{\Am}{{\rm{\bf A}}^{\!-} }
\newcommand{\Ap}{{\rm{\bf A}}^{\!+} }
\newcommand{\Apm}{{\rm{\bf A}}^{\!\pm} }

\newcommand{\h}{{\bf h}}

\newcommand{\hf}{{\scriptstyle \frac{1}{2}}}

\newcommand{\alb}{\boldsymbol{\alpha}}
\newcommand{\beb}{\boldsymbol{\beta}}

\newcommand{\bt}{\mathfrak{b}}

\begin{document}

\title[Spin chains from Onsager algebras]
{Quantum spin chains from Onsager algebras
\\ and reflection $K$-matrices}

\author{Atsuo Kuniba}
\address{Atsuo Kuniba: Institute of Physics, 
University of Tokyo, Komaba, Tokyo 153-8902, Japan}
\email{atsuo.s.kuniba@gmail.com}

\author{Vincent Pasquier}
\address{Vincent Pasquier: Institut de Physique Th\'eorique, 
Universit\'e Paris Saclay, CEA, CNRS, F-91191 
Gif-sur-Yvette, France}
\email{vincent.pasquier@ipht.fr}

\maketitle

\vspace{0.5cm}
\begin{center}{\bf Abstract}
\end{center}

We present a representation of the 
generalized $p$-Onsager algebras
$O_p(A^{(1)}_{n-1})$,
$O_p(D^{(2)}_{n+1})$, 
$O_p(B^{(1)}_n)$,
$O_p(\tilde{B}^{(1)}_n)$ and
$O_p(D^{(1)}_n)$ 
in which the generators are expressed as  
local Hamiltonians of XXZ type spin chains
with various boundary terms reflecting the Dynkin diagrams.
Their symmetry is described by the 
reflection $K$ matrices which are obtained recently by   
a $q$-boson matrix product construction related to the 3D integrability 
and characterized by Onsager coideals of quantum affine algebras.
The spectral decomposition of the $K$ matrices with respect to the 
classical part of the Onsager algebra is described conjecturally.
We also include a proof of a certain invariance property of 
boundary vectors in the $q$-boson Fock space 
playing a key role in the matrix product construction.
\vspace{0.4cm}

\section{Introduction}

The generalized $p$-Onsager algebra $O_p(A^{(1)}_{n-1})$ is 
generated by $\bt_1,\ldots, \bt_n$ with the relations
\begin{equation}\label{iys}
\begin{split}
&\bt_i\bt_j - \bt_j\bt_i = 0 \quad(a_{ij}=0),
\\
&\bt_i^2\bt_j -(p^2+p^{-2})\bt_i\bt_j\bt_i+\bt_j\bt_i^2 = \bt_j \quad 
(a_{ij}=-1),
\end{split}
\end{equation}
where we assume $n \ge 3$ and the parameter $p$ is generic.
The data $(a_{ij})_{i,j \in \Z_n}$ is the Cartan matrix of
the affine Lie algebra $A^{(1)}_{n-1}$ \cite{Kac}.
The above relation with $p=1$ goes back to \cite[eqs.(11),(12)]{UI}.
In what follows, the algebra $O_p(\mathfrak{g})$ introduced for 
any affine Lie algebra $\mathfrak{g}$ \cite{BB}  
will simply be called an Onsager algebra for short.
We refer to \cite[Rem. 9.1]{T} for the early history 
of the Onsager algebra starting from \cite{On} and  
\cite[Sec.1(1)]{Ko} for an account on more recent studies 
and the references therein. 

Let $q$ be a parameter such that $q^2=-p^{-2}$.
Then $O_p(A^{(1)}_{n-1})$ has a representation $\bt_i \mapsto b_i$ defined by 
\begin{align}\label{kuwako}
b_i  = z^{-\delta_{i,0}}\sigma^+_i\sigma^-_{i+1}
+ z^{\delta_{i,0}}\sigma^-_i\sigma^+_{i+1}
+\frac{q+q^{-1}}{4}\sigma^z_i\sigma^z_{i+1}
+\frac{q-q^{-1}}{4}(\sigma^z_i-\sigma^z_{i+1})
-\frac{(q-q^{-1})^2}{4(q+q^{-1})}
\quad (i \in \Z_n),
\end{align}
where $z$ is a spectral parameter and 
$\sigma^\pm_i= \frac{1}{2}(\sigma^x_i \pm \I \sigma^y_i), \sigma^z_i$ 
are the Pauli matrices acting on the
$i$ th component of $(\C^2) ^{\otimes n}$.
Thus the generators of 
$O_p(A^{(1)}_{n-1})$ are expressed as
local Hamiltonians of XXZ type spin chain on a length $n$ periodic lattice.

In this paper we explain the origin of 
the representation (\ref{kuwako}) and 
extend it to the Onsager algebra 
$O_p(\mathfrak{g})$ associated with the
non-exceptional affine Lie algebras\footnote{
$\tilde{B}^{(1)}_n$ is isomorphic to $B^{(1)}_n$ but
with a different numeration of the vertices of the Dynkin diagram.
See Section \ref{sec:bt}.
It is included for uniformity of the description.}   
$\mathfrak{g} = D^{(2)}_{n+2}$, $B^{(1)}_n$, 
$\tilde{B}^{(1)}_n$ and $D^{(1)}_n$.
It is based on the two recent works;
the $q$-boson matrix product construction\footnote{
The coexistence of $p$ and $q$ related by $p=\pm \I q^{-1}$ 
originates in the $q$-boson for $U_p$ in \cite{KP}.} 
of the reflection $K$ matrices
connected to the three dimensional (3D) integrability \cite{KP}
and the characterization of those $K$ matrices by 
{\em Onsager coideals} \cite{KOY2}. 
The resulting Hamiltonians contain various boundary terms 
reflecting the shape of the relevant Dynkin diagrams.
They yield a systematic realization of the Onsager algebras 
by spin chain Hamiltonians,
providing examples beyond free-fermions which were 
sought eagerly in the very end of \cite{UI}.

Let us sketch some more detail of our approach and the results. 
Our representation of $O_p(\mathfrak{g})$ is an elementary 
consequence of the composition
\begin{equation}\label{cdkmoyaru}
\begin{split}
&O_p(\mathfrak{g}) \hookrightarrow 
U_p(\mathfrak{g}) \rightarrow \mathrm{End} V
\\
& \quad\; \bt_i\; \longmapsto \,\;g_i \;\; \longmapsto\;\; b_i,
\end{split}
\end{equation}
where $U_p(\mathfrak{g})$ denotes the 
Drinfeld-Jimbo quantum affine algebra \cite{D,Ji}.
The space $V$ is taken as $(\C^2)^{\otimes n}$
and the latter arrow stands for the 
{\em fundamental representations} for $U_p(A^{(1)}_{n-1})$ 
and the {\em spin representations} 
for $U_p(\mathfrak{g})$ for the other $\mathfrak{g}$ mentioned in the above.
They carry a spectral parameter $z$ whose dependence is incorporated 
into $b_0$ only. (See Remark \ref{re:aoda} however.)
A natural basis of $V$ is parametrized as
$|\alpha_1,\ldots, \alpha_n \rangle$ 
with $\alpha_i \in \{0,1\}$, which may be viewed as a state of 
a spin ${\textstyle \frac{1}{2}}$ chain with $n$ sites.
In this interpretation, the generators 
$e_i, f_i, k^{\pm 1}_i$ of $U_p(\mathfrak{g})$ are expressed as  
exchange type interactions among local spins around site $i$ of the lattice.
See for instance 
(\ref{ayam})--(\ref{shnk}) for the typical examples in $U_p(A^{(1)}_{n-1})$
and also (\ref{hota1}), (\ref{hota2}) for peculiar ones in 
$U_p(D^{(1)}_n)$ involving ``pair creation/annihilation" 
of two boundary spins.
A crux here is to accommodate the length $n$ chain within 
a {\em single} $U_p(\mathfrak{g})$ module $V$
rather than considering the 
$n$-fold tensor product of the spin ${\textstyle \frac{1}{2}}$ 
representation of $U_p(\widehat{sl}_2)$.
Such an approach to size $n$ systems by rank $n$ algebras
has also turn out efficient in the mutispecies TASEP \cite{KMO}.

The first arrow in (\ref{cdkmoyaru}) stands for an algebra homomorphism
defined by 
\begin{align}\label{bgi}
\bt_i \mapsto g_i = f_i + p^2 k^{-1}_i e_i + \frac{1}{q+q^{-1}}k^{-1}_i
\quad (i \in \Z_n)
\end{align}  
for $\mathfrak{g}=A^{(1)}_{n-1}$.
A similar embedding is known for all $\mathfrak{g}$ \cite{BB}.
In this context, (\ref{iys}) can be viewed as a modified 
$p$-Serre relation.
Observe in general that the elements of the form 
$g'_i = f_i + c_i k^{-1}_i e_i + d_i k^{-1}_i \in U_p(\mathfrak{g})$
with arbitrary coefficients $c_i, d_i$ behave under the coproduct 
$\Delta$ (defined in (\ref{Del})) as
\begin{equation*}
\begin{split}
\Delta g'_i &= f_i \otimes 1 + k^{-1}_i \otimes f_i 
+ c_i (k^{-1}_i \otimes k^{-1}_i)(1\otimes e_i + e_i \otimes k_i)
+ d_i  k^{-1}_i \otimes k^{-1}_i
\\
&= (f_i + c_i k^{-1}_i e_i) \otimes 1 + k^{-1}_i \otimes g'_i
\in U_p \otimes 1 + U_p \otimes g'_i.
\end{split}
\end{equation*}
It implies that the subalgebra $\mathcal{B}_p \subset U_p$ 
generated by $g'_i$'s becomes a left coideal subalgebra 
$\Delta \mathcal{B}_p \subset U_p \otimes \mathcal{B}_p$.
In this vein, 
the coideal subalgebra of $U_p(A^{(1)}_{n-1})$ generated by $g_i$ (\ref{bgi}) 
whose coefficients are deliberately chosen to further fit (\ref{iys}) 
was called an Onsager coideal in \cite{KOY2}.
Its natural analogue for $\mathfrak{g}$ other than $A^{(1)}_{n-1}$ 
can also be formulated,
albeit that a couple of variants are allowed for the coefficients $c_i, d_i$.
See (\ref{bg0})--(\ref{drk}) and the remarks following them.

Having the Onsager coideals of $U_p(\mathfrak{g})$ of a decent origin, 
it is tempting to seek the associated {\em reflection $K$ matrices} governed by them 
via the boundary intertwining relation \cite{DM}. 
In our setting it is represented as the symmetry of local Hamiltonians 
\begin{align}\label{aodai}
(b_i|_{z \rightarrow z^{-1}}) K(z) = K(z) b_i\quad (0 \le i \le n'),
\end{align}
where the replacement $z \rightarrow z^{-1}$ is relevant only for $i=0$.
The integer $n'$ denotes the rank of $\mathfrak{g}$, i.e., 
$n'=n-1$ for $\mathfrak{g} = A^{(1)}_{n-1}$ and $n$ for the other type
under consideration.
It was shown in \cite{KOY2} that (\ref{aodai}) 
admits a unique (up to normalization) solution $K(z): V \rightarrow V$
satisfying the reflection equation \cite{Ch,Kul,Sk}.
Moreover it reproduces the reflection $K$ matrix 
constructed by the matrix product method connected to the
3D integrability \cite{KP}.
In other words, these $K$ matrices are characterized by 
the commutativity with the local Hamiltonians.

Introduce the Hamiltonian 
$H(z) = \kappa_0 b_0 + \kappa_1 b_1 + \cdots + \kappa_{n'}b_{n'}$
with constant coefficients $\kappa_0,\ldots, \kappa_{n'}$.
It depends on $z$ via $b_0$ only.
Then (\ref{aodai}) implies the quasi-commutativity 
$H(z^{-1})K(z) = K(z) H(z)$ for {\em arbitrary} 
$\kappa_0,\ldots, \kappa_{n'}$.
In the special case $\kappa_0=0$,
$H(z)$ reduces to 
a $z$-independent operator $H$ enjoying the symmetry $[K(z),H]=0$.
On the other hand, for each $\mathfrak{g}$ under consideration, 
we will show that there is one special choice of 
$\kappa_0,\ldots, \kappa_{n'}$ (up to overall normalization) 
such that all of them are non-vanishing and 
\begin{align*}
[K^\vee(z),H(z)]=0.
\end{align*}
Here $K^\vee(z)=\sigma^x K(z)$ and 
$\sigma^x$ is the global spin reversal operator (\ref{sx}).
In this way, the $K$ matrices in \cite{KP} are shown to
serve as various versions of symmetry operators of 
the Hamiltonians consisting of Onsager algebra generators. 
See also the ending remarks in Section \ref{sec:sum}.

Our second main result is the spectral decomposition of the 
$K$ matrices with respect to the {\em classical part} 
$O_p(\overline{\mathfrak{g}})$ of the 
Onsager algebra $O_p(\mathfrak{g})$\footnote{
$\;\overline{\mathfrak{g}}$ is obtained from $\mathfrak{g}$ 
by removing the 0 th vertex in its Dynkin diagram.}.
The former is defined as the subalgebra of the latter by
dropping the generator $\bt_0$.
The relation (\ref{aodai}) tells that 
$K(z)$ commutes with $O_p(\overline{\mathfrak{g}})$ in the 
representation $V$ under consideration.
Therefore it is a scalar on each irreducible 
$O_p(\overline{\mathfrak{g}})$ component within $V$.
We present detailed conjectures on 
the eigenspectra  and the decompositions.
A typical formula of such kind is (\ref{sept}).
They are boundary analogues of the celebrated spectral decomposition
of quantum $R$ matrices with respect to $U_p(\overline{\mathfrak{g}})$,
and deserve further studies from the viewpoint of representation theory of
Onsager algebras.

Our third main result is a proof of Theorem \ref{th:bv} in Appendix \ref{app:A}.
It states certain vectors in the $q$-boson Fock space remain invariant 
under the action of the intertwiner of the quantized coordinate ring 
$A_q(\mathrm{Sp}_4)$ \cite{KO}.
The content is apparently independent from the other parts of the paper.
However the claim is essential and has been 
used as a key conjecture in \cite{KP} to perform 
the $q$-boson matrix product construction of the  
$K$ matrices for $\mathfrak{g} = D^{(2)}_{n+2}$, $B^{(1)}_n$, 
$\tilde{B}^{(1)}_n$ and $D^{(1)}_n$ treated in this paper.
So the proof included here really completes the 3D approach by the 
authors \cite{KP}
and establishes the reflection equation independently from the   
representation theoretical method using Onsager coideals \cite{KOY2}.

The layout of the paper is as follows.
In Section \ref{sec:cdko},
quantum affine algebras $U_p(\mathfrak{g})$ and 
the $q$-boson matrix product construction of the reflection $K$ matrices \cite{KP}
are recalled.
In Section \ref{sec:A}, fundamental representations of $U_p(A^{(1)}_{n-1})$
are recalled and the Hamiltonian associated with $O_p(A^{(1)}_{n-1})$ is given.
A simple connection to the Temperley-Lieb algebra \cite{TL} is pointed out
in Remark \ref{re:tl}.
In Section \ref{sec:spda}, spectral decomposition of the type $A^{(1)}_{n-1}$ 
$K$ matrix with respect to the classical part of $O_p(A^{(1)}_{n-1})$
is described.
Section \ref{sec:tatsuki} is a guide to the 
subsequent sections devoted to presenting parallel results for
$O_p(\mathfrak{g})$ with 
$\mathfrak{g}$ other than $A^{(1)}_{n-1}$.
It summarizes common and general features in these cases.
Concrete formulas for the spin representations of $U_p(\mathfrak{g})$,  
Hamiltonians associated with $O_p(\mathfrak{g})$ 
and their $K$ matrix symmetry are given in 
Section \ref{sec:D2} for $D^{(2)}_{n+1}$,
Section \ref{sec:B} for $B^{(1)}_{n}$,
Section \ref{sec:bt} for $\tilde{B}^{(1)}_{n}$
and
Section \ref{sec:D1} for $D^{(1)}_{n}$.
Section \ref{sec:spdb} and Section \ref{sec:spdd} 
describe the spectral decompositions of the $K$ matrices 
when the classical part of $O_p(\mathfrak{g})$ is
$O_p(B_n)$ and $O_p(D_n)$, respectively.
Section \ref{sec:sum} is a summary. 
Appendix \ref{app:kcom} is a proof of commutativity 
of the $K$ matrix for type $A^{(1)}_{n-1}$.
Appendix \ref{app:A} contains a proof of the important 
Theorem \ref{th:bv}.
Throughout the paper the parameters $q, p$ are 
related by (\ref{pa1}) and assumed to be generic.
We use the notation
\begin{align*}
(z;q)_m = \prod_{j=1}^m(1-zq^{j-1}).
\end{align*}

\section{General remarks and definitions}\label{sec:cdko}
In this section we 
introduce the definitions that will be commonly used in the paper.

\subsection{Quantum affine algebra $U_p$}\label{ss:qaa}
Let 
$U_p=U_p(A^{(1)}_{n-1})\,(n\ge 3)$, 
$U_p(D^{(2)}_{n+1})\, (n \ge 2)$, 
$U_p(B^{(1)}_n)\,(n\ge 3)$,
$U_p(\tilde{B}^{(1)}_n)\,(n\ge 3)$, 
$U_p(D^{(1)}_n)\, (n\ge 3)$
be quantum affine algebras without derivation operator \cite{D,Ji}. 
The affine Lie algebra $\tilde{B}^{(1)}_n$ is just 
$B^{(1)}_n$ but with different enumeration of the vertices 
as shown in Section \ref{ss:opbt}.
Note that $U_p(A^{(1)}_{1})$ has been excluded.
We assume that $p$ is generic throughout.
For convenience set 
$n'= n-1$ for $A^{(1)}_{n-1}$ and 
$n'=n$ for the other cases.
$U_p$ is a Hopf algebra 
generated by $e_i, f_i, k^{\pm 1}_i\, (0 \le i \le n')$ satisfying 
\begin{equation}\label{uqdef}
\begin{split}
&k_i k^{-1}_i = k^{-1}_i k_i = 1,\quad k_i k_j=k_jk_i,\\
&k_ie_jk^{-1}_i = p_i^{a_{ij}}e_j,\quad 
k_if_jk^{-1}_i = p_i^{-a_{ij}}f_j,\quad
e_i f_j-f_je_i=\delta_{i,j}\frac{k_i-k^{-1}_i}{p_i-p^{-1}_i},
\end{split}
\end{equation}
and the Serre relations which will be described later.
The Cartan matrix $(a_{ij})_{0 \le i,j \le n'}$ \cite{Kac}
will also be given later for each case.
The constants $p_i\, (0 \le i \le n')$ in (\ref{uqdef}) are 
$p_i=p^2$ except for
$p_0 = p_n = p$ for $D^{(2)}_{n+1}$,
$p_n = p$ for $B^{(1)}_{n}$ and
$p_0 = p$ for $\tilde{B}^{(1)}_{n}$.
In addition to $p$, we allow the 
coexistence of the parameters $q, t$ and the sign factors 
$\epsilon, \mu$ related as
\begin{align}\label{pa1}
q^{\frac{1}{2}} = \I \mu t,
\quad 
p= -\I \epsilon q^{-1} = \I \epsilon t^{-2},\quad \quad  
\epsilon=\pm 1, \quad \mu = \pm 1.
\end{align} 
The second relation is the same with \cite[eq.(96)]{KOY2}.
The coproduct $\Delta$ is taken as 
\begin{align}\label{Del}
\Delta k^{\pm 1}_i = k^{\pm 1}_i\otimes k^{\pm 1}_i,\quad
\Delta e_i = 1\otimes e_i + e_i \otimes k_i,\quad
\Delta f_i = f_i\otimes 1 + k^{-1}_i\otimes f_i.
\end{align}

\subsection{$U_p$ module $V$ and local spins}

We will be concerned with the $U_p$ module $V$ with 
$\dim V = 2^n$ presented as
\begin{align}
V &=  \bigoplus_{\alb \in \{0,1\}^n}\C |\alb\rangle\simeq (\C^2)^{\otimes n},
\label{Vdef}
\\
|\alb\rangle & = |\alpha_1, \ldots, \alpha_n\rangle,\quad
|\alb| = \alpha_1+\cdots + \alpha_n 
\;\;\;\text{for}\;\; \alb = (\alpha_1, \ldots, \alpha_n),
\quad \alpha_i \in \{ 0,1\}.
\label{aoda}
\end{align}
Vectors $|\beb\rangle$ with $\beb \not\in \{0,1\}^n$ 
should be understood as $0$.
The space $V$ will be an irreducible $U_p$ module 
for $U_p(D^{(1)}_{n+1})$, 
$U_p(B^{(1)}_{n})$ and 
$U_p(\tilde{B}^{(1)}_{n})$.
For $U_p(A^{(1)}_{n-1})$ and $U_p(D^{(1)}_n)$, one needs to introduce 
the finer subspaces $V_l$ and $V_\pm$ as
\begin{align}
V_l = \bigoplus_{\alb \in \{0,1\}^n, \, |\alb | = l}\C |\alb\rangle,
\qquad 
V_\pm = \bigoplus_{\alb \in \{0,1\}^n,\, (-1)^{|\alb |} = \pm 1}\C |\alb\rangle,
\label{vl}
\end{align}
which leads to the decompositions 
\begin{align}
V = V_0 \oplus V_1 \oplus \cdots \oplus V_n,\qquad
V = V_+ \oplus V_-.
\end{align} 
The explicit $U_p$ module structure of $V$ 
will be described in the subsequent sections.
They have appeared for example in \cite[Sec.2]{DO}
for $U_p(A^{(1)}_{n-1})$, and in 
\cite[Sec.B.2]{KP} for the other types.

Let $\sigma^{x}_i, \sigma^{y}_i, \sigma^{z}_i$ and 
$\sigma^\pm_i = \frac{1}{2}(\sigma^x_i \pm \I \sigma^y_i)$ $(1 \le i \le n)$,
denote the Pauli matrices acting on the $i$ th component of $V$
regarding $\alpha_i=1$ as an up-spin and 
$\alpha_i=0$ as a down-spin. Namely, 
\begin{alignat}{2}
\sigma^x_i|\ldots, 1,\ldots\rangle &= |\ldots, 0,\ldots\rangle,
&\qquad
\sigma^x_i|\ldots, 0,\ldots\rangle &= |\ldots, 1,\ldots\rangle,
\nonumber\\
\sigma^y_i|\ldots, 1,\ldots\rangle &= \I |\ldots, 0,\ldots\rangle,
&\qquad
\sigma^y_i|\ldots, 0,\ldots\rangle &= -\I|\ldots, 1,\ldots\rangle,
\nonumber\\
\sigma^z_i|\ldots, 1,\ldots\rangle &= |\ldots, 1,\ldots\rangle,
&\qquad
\sigma^z_i|\ldots, 0,\ldots\rangle &= -|\ldots, 0,\ldots\rangle,
\label{ls}\\
\sigma^+_i|\ldots, 1,\ldots\rangle &= 0,
&\qquad
\sigma^+_i|\ldots, 0,\ldots\rangle &= |\ldots, 1,\ldots\rangle,
\nonumber\\
\sigma^-_i|\ldots, 1,\ldots\rangle &= |\ldots, 0,\ldots\rangle,
&\qquad
\sigma^-_i|\ldots, 0,\ldots\rangle &= 0.
\nonumber
\end{alignat}
The global spin reversal operator will be denoted by 
\begin{align}\label{sx}
\sigma^x = \sigma^x_1\sigma^x_2 \cdots \sigma^x_n.
\end{align}
It acts on a base vector as
$\sigma^x|\alb\rangle = |{\bf 1}-\alb\rangle$
where ${\bf 1}= {\bf e}_1+\cdots + {\bf e}_n$.

\subsection{$K$ matrices}\label{ss:kmat}
Let us recall the matrix product construction of the $K$ matrices 
related to the 3D integrability \cite{KP}.
We will not use the reflection equations satisfied by them in this paper. 
They have been described in detail in \cite{KP,KOY2}.

Let 
$F_q = \bigoplus_{m\ge 0}\C |m\rangle$ 
and $F^\ast_q = \bigoplus_{m \ge 0} \C\langle m |$ be 
the Fock space and its dual 
equipped with the inner product 
$\langle m | m'\rangle = (q^2;q^2)_m\delta_{m,m'}$.
We define the $q$-boson operators $\ap, \am, \ok$ on them by 
\begin{equation*}
\begin{split}
&\ap |m\rangle = |m+1\rangle,\quad
\am |m\rangle = (1-q^{2m})|m-1\rangle,\quad
\ok |m\rangle = q^{m} |m\rangle,\\
&\langle m | \am = \langle m+1 |, \quad
\langle m | \ap = \langle m-1| (1-q^{2m}),\quad
\langle m | \ok = \langle m| q^{m}.
\end{split}
\end{equation*}
They satisfy $(\langle m | X)|m'\rangle 
= \langle m | (X|m'\rangle)$ and the relations
\begin{align}\label{qbos}
\ok \,\apm = q^{\pm 1} \apm \ok,
\quad 
\ap \am = 1 - \ok^2,\quad
\am \ap = 1 - q^2 \ok^2.
\end{align}
We also use the number operator ${\bf h}$ acting as 
${\bf h} |m \rangle = m|m\rangle$ and 
$\langle m | {\bf h} = \langle m | m$ so that 
$\ok$ may be identified with  $q^{\bf h}$. 
Set 
\begin{align}\label{kop}
\begin{pmatrix}
K^0_0 & K^0_1 \\ K^1_0 & K^1_1
\end{pmatrix} = 
\begin{pmatrix}
\ap & -q\ok \\ \ok  & \am
\end{pmatrix}.
\end{align}

The $K$ matrix $K_{\mathrm{tr}}(z)$ related to 
$U_p(A^{(1)}_{n-1})$ is given by the matrix product formula \cite{KP}:
\begin{align}
K_{\mathrm{tr}}(z):&\; V \rightarrow V,
\quad
K_{\mathrm{tr}}(z)|\alb\rangle 
= \sum_{\beb \in \{0,1\}^n}
K_{\mathrm{tr}}(z)_{\alb}^{\beb} |\beb \rangle,
\label{ktr1}\\
K_{\mathrm{tr}}(z)_{\alb}^{\beb}& = 
\kappa_{\mathrm{tr},|\alb| }(z)
\mathrm{Tr}(z^{\bf h} K_{\alpha_1}^{\beta_1} \cdots 
K_{\alpha_n}^{\beta_n} ),
\label{mpktr}\\
\kappa_{\mathrm{tr},l}(z) &= (-1)^lq^{\min(0,2l-n)}
(1-q^{|n-2l|}z).
\label{kapa}
\end{align}
The trace here is evaluated by means of (\ref{qbos}) and 
$\mathrm{Tr}(z^{\bf h}\ok^m) = \frac{1}{1-zq^m}$.
All the elements $K_{\mathrm{tr}}(z)_{\alb}^{\beb}$ is a 
rational function of $z$ and $q$.
Moreover it is easily seen that 
$K_{\mathrm{tr}}(z)_{\alb}^{\beb}=0$ 
unless $|\alb| + |\beb|=n$.
Thus (\ref{ktr1}) is actually refined as
\begin{align}
K_{\mathrm{tr}}(z)&= K_{\mathrm{tr},0}(z)\oplus \cdots 
\oplus K_{\mathrm{tr},n}(z),
\qquad 
K_{\mathrm{tr},l}(z): \, V_l \rightarrow V_{n-l},
\label{ktrd}\\
K_{\mathrm{tr}}(z)|\alb\rangle 
&= \sum_{\beb \in \{0,1\}^n, |\beb|=n-|\alb|}
K_{\mathrm{tr}}(z)_{\alb}^{\beb} |\beb \rangle.
\end{align}
Some examples from $n=3$ read
\begin{align*}
K_{\mathrm{tr}}(z)|011\rangle
&= -\frac{q(1-q^2)z}{1-q^3z}|001\rangle
-\frac{q^2(1-q^2)z}{1-q^3z}|010\rangle
+\frac{q^2(1-qz)}{1-q^3z}|100\rangle,
\\
K_{\mathrm{tr}}(z)|101\rangle
&= -\frac{q^2(1-q^2)z}{1-q^3z}|001\rangle
+\frac{q^2(1-qz)}{1-q^3z}|010\rangle
-\frac{q(1-q^2)}{1-q^3z}|100\rangle,
\\
K_{\mathrm{tr}}(z)|110\rangle
&= \frac{q^2(1-qz)}{1-q^3z}|001\rangle
-\frac{q(1-q^2)}{1-q^3z}|010\rangle
-\frac{q^2(1-q^2)}{1-q^3z}|100\rangle,
\end{align*}
which are actually the action of $K_{\mathrm{tr},2}(z)$.
We have slightly changed the gauge in (\ref{kop}) from 
\cite[eq.(6)]{KP} and 
the normalization factor from \cite[eq.(77)]{KP}
to (\ref{kapa}) so that
\begin{align}\label{petb}
K_{\mathrm{tr}}(z)K_{\mathrm{tr}}(z^{-1})= \mathrm{id}_V
\end{align}
is satisfied.
Another notable property is the commutativity.
\begin{proposition}\label{pr:kk}
\begin{align}\label{com}
[K_{\mathrm{tr}}(z),K_{\mathrm{tr}}(w)]=0,
\end{align}
where $[\;,\;]$ denotes the commutator defined after (\ref{bbre3}).
\end{proposition}
A proof of  Proposition \ref{pr:kk} is given in Appendix \ref{app:kcom}.

To present the $K$ matrices $K_{k,k'}(z)$ related to 
$U_p(D^{(2)}_{n+1})$, 
$U_p(B^{(1)}_n)$,
$U_p(\tilde{B}^{(1)}_n)$, 
$U_p(D^{(1)}_n)$,
we prepare the {\em boundary vectors}
\begin{align}
\langle\eta_k| = \sum_{m\ge 0}\frac{\langle km|}{(q^{k^2};q^{k^2})_m} \in F^\ast_q,
\qquad
|\eta_k\rangle = \sum_{m\ge 0}\frac{|km\rangle}{(q^{k^2};q^{k^2})_m} \in F_q
\qquad (k=1,2).
\label{xb}
\end{align}
Then $K_{k,k'}(z)$ are given by the matrix product construction \cite{KP}:
\begin{align}
K_{k,k'}(z) :& \;V \rightarrow V, \quad 
K_{k,k'}(z) |\alb\rangle  = 
\sum_{\beb \in \{0,1\}^n}K_{k,k'}(z) _{\alb}^{\beb} |\beb\rangle
\quad (k,k'=1,2),
\label{kkk1}\\
K_{k,k'}(z) _{\alb}^{\beb} &=
\kappa_{k,k'}(z)
\langle \eta_k | z^{\bf h} K_{\alpha_1}^{\beta_1} \cdots 
K_{\alpha_n}^{\beta_n}  | \eta_{k'}\rangle
\quad ((k,k')\neq (2,2)),
\label{kkk2}\\
K_{2,2}(z) _{\alb}^{\beb} &=
\kappa_{2,2}(z)^{(-1)^n}
\langle \eta_2 | z^{\bf h} K_{\alpha_1}^{\beta_1} \cdots 
K_{\alpha_n}^{\beta_n}  | \eta_2\rangle,
\label{kkk3}
\end{align}
where the normalization factors are specified as 
\begin{align}
\kappa_{k,k'}(z) =\langle \eta_k | z^{\bf h}|\eta_{k'}\rangle^{-1}
= \frac{(z^{\max(k,k')};q^{kk'})_\infty}
{((-q)^{\min(k,k')}z^{\max(k,k')};q^{kk'})_\infty}.
\end{align}
The quantity 
$\langle \eta_k | z^{\bf h}X|\eta_{k'}\rangle$ 
for any polynomial $X$ in $\apm, \ok$ can be 
calculated by using (\ref{qbos}) and 
the explicit formula 
\begin{equation}\label{lin}
\begin{split}
&\langle \eta_k|z^{\bf h} (\apm)^j \ok^m 
w^{\bf h}|\eta_{k'}\rangle =
\langle \eta_{k'}|w^{\bf h}\ok^m  (\amp)^j 
z^{\bf h}|\eta_k\rangle\quad (k, k' =1,2),
\\
&\langle \eta_1|z^{\bf h}(\ap)^j \ok^m 
w^{\bf h}|\eta_1\rangle
= z^j(-q;q)_j
\frac{(-q^{j+m+1}zw;q)_\infty}{(q^mzw;q)_\infty},
\\
&\langle \eta_1|z^{\bf h}(\am)^j \ok^m 
w^{\bf h}|\eta_2\rangle
= z^{-j}\sum_{i=0}^j(-1)^i q^{\frac{1}{2}i(i+1-2j)}
\frac{(q;q)_j}{(q;q)_i(q;q)_{j-i}}
\frac{(-q^{2i+2m+1}z^2w^2;q^2)_\infty}{(q^{2i+2m}z^2w^2;q^2)_\infty},
\\
&\langle \eta_1|z^{\bf h} (\ap)^j \ok^m 
w^{\bf h}|\eta_2\rangle
= z^{j}\sum_{i=0}^j q^{\frac{1}{2}i(i+1)}
\frac{(q;q)_j}{(q;q)_i(q;q)_{j-i}}
\frac{(-q^{2i+2m+1}z^2w^2;q^2)_\infty}{(q^{2i+2m}z^2w^2;q^2)_\infty},
\\
&\langle \eta_2|z^{\bf h} (\ap)^{j} \ok^m 
w^{\bf h}|\eta_2\rangle
= \theta(j\in 2\Z) \,z^{j}(q^2;q^4)_{\frac{j}{2}}
\frac{(q^{2j+2m+2}z^2w^2;q^4)_\infty}{(q^{2m}z^2w^2;q^4)_\infty},
\end{split}
\end{equation}
where $\theta(\text{true})=1, \theta(\text{false})=0$.
Obviously $K_{2,2}(z)_{\alb}^{\beb} = 0$ unless 
$|\alb| +|\beb|-n \in 2\Z$.
Therefore  (\ref{kkk1}) for $(k,k') = (2,2)$ is refined as
\begin{align}
K_{2,2}(z) &= K_{2,2,+}(z) \oplus K_{2,2,-}(z),
\qquad
K_{2,2,\pm}(z): V_\pm \rightarrow V_{\pm (-1)^n},
\label{askB}\\
K_{2,2}(z)|\alb\rangle &= 
\sum_{\beb \in \{0,1\}^n, |\beb| \in n-|\alb|+ 2\Z} 
K_{2,2}(z)_{\alb}^{\beb} |\beb\rangle.
\end{align}
The normalization factors have been chosen so that
all the elements of $K_{k,k'}(z)$ are rational function of $z,q$ and
\begin{align}
K_{k,k'}(z)_{0,0, \ldots, 0}^{1,1, \ldots, 1}
&= \frac{(z^{\max(k,k')};q^{kk'})_n}
{(-qz^{\max(k,k')};q^{kk'})_n}\quad ((k,k')\neq (2,2)),
\\
K_{2,2}(z)_{0,0, \ldots, 0}^{1,1, \ldots, 1}
&= -q^{-1}K_{2,2}(z)_{1,0, \ldots, 0}^{0,1, \ldots, 1}
= \theta(n\in 2\Z) \frac{(z^2;q^4)_{\frac{n}{2}}}{(q^2z^2;q^4)_{\frac{n}{2}}}
+\theta(n\in 2\Z+1)
\frac{(q^2z^2;q^4)_{\frac{n-1}{2}}}{(z^2;q^4)_{\frac{n+1}{2}}}.
\end{align}
For instance one has
\begin{align*}
K_{1,1}(z)|00\rangle 
& = \frac{(-q;q)_2z^2|00\rangle}{(-qz;q)_2}
+\frac{(1+q)(1-z)z|01\rangle}{(-qz;q)_2}
+\frac{q(1+q)(1-z)z|10\rangle}{(-qz;q)_2}
+\frac{(z;q)_2|11\rangle}{(-qz;q)_2},
\\
K_{1,2}(z)|00\rangle &= 
\frac{(1+q)z^2(1+q^2-q^2z^2+q^3z^2)|00\rangle}
{(-qz^2;q^2)_2}
+\frac{(1+q)(1-z^2)z|01\rangle}{(-qz^2;q^2)_2}\\
&+\frac{q(1+q)(1-z^2)z|10\rangle}{(-qz^2;q^2)_2}
+\frac{(z^2;q^2)_2|11\rangle}{(-qz^2;q^2)_2},
\\
K_{2,1}(z)|00\rangle &= 
\frac{(1+q)z^2(1-q+qz^2+q^3z^2)|00\rangle}
{(-qz^2;q^2)_2}
+\frac{q(1+q)(1-z^2)z^2|01\rangle}{(-qz^2;q^2)_2}\\
&+\frac{q^2(1+q)(1-z^2)z^2|10\rangle}{(-qz^2;q^2)_2}
+\frac{(z^2;q^2)_2|11\rangle}{(-qz^2;q^2)_2},
\\
K_{2,2}(z)|000\rangle &= 
\frac{(1-q^2)z^2|001\rangle}{(z^2;q^4)_2}
+\frac{q(1-q^2)z^2|010\rangle}{(z^2;q^4)_2}
+\frac{q^2(1-q^2)z^2|100\rangle}{(z^2;q^4)_2}
+\frac{(1-q^2z^2)|111\rangle}{(z^2;q^4)_2},
\\
K_{2,2}(z)|100\rangle&=
-\frac{q^3(1-q^2)z^2|000\rangle}{(z^2;q^4)_2}
-\frac{q(1-q^2z^2)|011\rangle}{(z^2;q^4)_2}
+\frac{(1-q^2)|101\rangle}{(z^2;q^4)_2}
+\frac{q(1-q^2)|110\rangle}{(z^2;q^4)_2}.
\end{align*}
The commutativity (\ref{com}) does not hold for the 
$K$ matrices $K_{k,k'}(z)$. 

For later convenience 
let us introduce two slight variants of the $K$ matrices.
The first one is a gauge transformation of $K_{k,k'}(z)$ as
\begin{align}
\tilde{K}_{k,k'}(z) & = S K_{k,k'}(z) S^{-1}\quad (k,k'=1,2),
\label{ktd}
\\
S &= \mathrm{diag}(S_{\alb}) _{\alb \in \{0,1\}^n},
\quad
S_{\alb} = (\I q^{\frac{1}{2}})^{|\alb|}
= (-\mu t)^{|\alb|}.
\end{align}
See (\ref{pa1}) for the relation among the parameters $q$ and $t$.
It is symmetric, i.e., 
\begin{align}
\tilde{K}_{k,k'}(z)|\alb\rangle = \sum_{\beb \in \{0,1\}^n}
\tilde{K}_{k,k'}(z)_{\alb}^{\beb}|\beb\rangle,
\qquad
\tilde{K}_{k,k'}(z)_{\alb}^{\beb} = 
\tilde{K}_{k,k'}(z)^{\alb}_{\beb}.
\end{align}
In fact, the elements 
$\tilde{K}_{k,k'}(z)_{\alb}^{\beb} $ are obtained from
(\ref{kkk2}) and (\ref{kkk3}) by 
replacing the local matrix product operators (\ref{kop}) 
by a symmetrized one:
\begin{align}\label{kops}
\begin{pmatrix}
\tilde{K}^0_0 & \tilde{K}^0_1 \\ \tilde{K}^1_0 & \tilde{K}^1_1
\end{pmatrix} = 
\begin{pmatrix}
1 & 0\\ 0 & \I q^{\frac{1}{2}}
\end{pmatrix}
\begin{pmatrix}
K^0_0 & K^0_1 \\ K^1_0 & K^1_1
\end{pmatrix}
\begin{pmatrix}
1 & 0\\ 0 & \I q^{\frac{1}{2}}
\end{pmatrix}^{-1}
=
\begin{pmatrix}
\ap & \I q^{\frac{1}{2}}\ok \\ \I q^{\frac{1}{2}}\ok  & \am
\end{pmatrix}.
\end{align}

The second variant of the $K$ matrices is defined by 
\begin{align}\label{kve}
K^\vee_{\mathrm{tr}}(z) = \sigma^x K_{\mathrm{tr}}(z),
\qquad
K^\vee_{k,k'}(z) = \sigma^x \tilde{K}_{k,k'}(z)
\quad (k,k'=1,2),
\end{align}
where $\sigma^x$ is the spin reversal operator (\ref{sx}). 
By the definition their matrix elements are related to the original ones just by
$K^\vee_{\mathrm{tr}}(z)_{\alb}^{\beb}
=K_{\mathrm{tr}}(z)_{\alb}^{{\bf 1}-\beb}$ and 
$K^\vee_{k,k'}(z)_{\alb}^{\beb}
=\tilde{K}_{k,k'}(z)_{\alb}^{{\bf 1}-\beb}$.
As with $K_{\mathrm{tr}}(z)$ and 
$K_{k,k'}(z)$, they are linear maps on $V$.
A notable contrast with (\ref{ktrd}) and (\ref{askB}) is that   
they now preserve the nontrivial subspaces which exist for  
$U_p(A^{(1)}_{n-1})$ and $U_p(D^{(1)}_n)$:
\begin{align}
K^\vee_{\mathrm{tr}}(z)&= K^\vee_{\mathrm{tr},0}(z)\oplus \cdots 
\oplus K^\vee_{\mathrm{tr},n}(z),
\qquad 
K^\vee_{\mathrm{tr},l}(z): \, V_l \rightarrow V_l,
\label{ktrd2}\\
K^\vee_{2,2}(z) &= K^\vee_{2,2,+}(z) \oplus K^\vee_{2,2,-}(z),
\qquad\;\;
K^\vee_{2,2,\pm}(z): V_\pm \rightarrow V_{\pm}.
\label{askB2}
\end{align}

\section{$O_p(A^{(1)}_{n-1})$ Hamiltonians}\label{sec:A}

We first present the results for $A^{(1)}_{n-1}$ case 
in this section and the next.

\subsection{$U_p(A^{(1)}_{n-1})$ and fundamental representations}

We assume $n \ge 3$.
The Dynkin diagram and the Cartan matrix are given by
\begin{align*}
\begin{picture}(200,50)
\put(30,0){
\put(0,24){$A^{(1)}_{n-1}$}
\drawline(20,3)(67,30)
\put(70,30){\circle{6}}
\drawline(73,30)(120,3)
\multiput( 20,0)(20,0){2}{\circle{6}}
\multiput(100,0)(20,0){2}{\circle{6}}
\multiput(23,0)(20,0){2}{\line(1,0){14}}
\put(83,0){\line(1,0){14}}\put(103,0){\line(1,0){14}}
\put(20,-6){\makebox(0,0)[t]{$1$}}
\put(40,-6){\makebox(0,0)[t]{$2$}}
\put(100,-6){\makebox(0,0)[t]{$n\!\! -\!\! 2$}}
\put(122,-6){\makebox(0,0)[t]{$n\!\! -\!\! 1$}}
\put(67,17){0}
}
\end{picture}
(a_{ij})_{0\le i,j \le n-1}= 
\begin{pmatrix}
2 & -1  & \cdots &  0 & -1
\\
-1 & 2  & \cdots  & & 0 
\\
\vdots   & \ddots & \ddots & \ddots &  \vdots \\
0 &  &  & 2 &  -1
\\
-1 & 0 & \cdots  & -1 & 2
\end{pmatrix}.
\end{align*}
The Serre relations have the form
\begin{align}
&e_ie_j - e_je_i = 0 \quad (a_{ij}=0),
\label{sr0}\\
&e_i^2e_j -(p^2+p^{-2})e_ie_je_i+e_je_i^2 = 0\quad (a_{ij}=-1),
\label{sr00}
\end{align}
and the same ones for $f_j$'s.
The fundamental representations are defined on the 
subspaces $V_0, V_1, \ldots, V_n$ of $V$ in (\ref{vl}) as
\begin{align}\label{a:re}
e_j|\alb \rangle 
= z^{\delta_{j,0}}|\alb-{\bf e}_j+{\bf e}_{j+1}\rangle, 
\quad
f_j|\alb \rangle 
= z^{-\delta_{j,0}}|\alb+{\bf e}_j-{\bf e}_{j+1}\rangle,
\quad
k_j |\alb \rangle
= p^{2(\alpha_{j+1}-\alpha_{j})}|\alb\rangle
\quad (j \in \Z_n),
\end{align}
where $z$ is a spectral parameter.
The symbol ${\bf e}_j$ denotes the $j$ th elementary vector
\begin{align}
{\bf e}_j = (0,\ldots, 0, \overset{j}{1},0,\ldots,0) \in \Z^n \;\;
(1 \le j \le n).
\end{align}
This should not be confused with the generator $e_j$ of $U_p$.

\subsection{Onsager algebra $O_p(A^{(1)}_{n-1})$ and the classical part 
$O_p(A_{n-1})$}\label{ss:oaa1}
Again we assume $n \ge 3$.
The algebra 
$O_p(A^{(1)}_{n-1})$ is generated by $\bt_0,\ldots, \bt_{n-1}$
obeying the relations \cite{BB}:
\begin{align}
&\bt_i\bt_j - \bt_j\bt_i = 0 \quad (a_{ij}=0),
\label{oar1}\\
&\bt_i^2\bt_j -(p^2+p^{-2})\bt_i\bt_j\bt_i+\bt_j\bt_i^2 = \bt_j \quad (a_{ij}=-1).
\label{oar2}
\end{align}
The classical part of $A^{(1)}_{n-1}$ without the vertex 0 is $A_{n-1}$.
Thus the subalgebra of $O_p(A^{(1)}_{n-1})$ generated by 
$\bt_1,\ldots, \bt_{n-1}$ is the Onsager algebra for $A_{n-1}$.
We denote it by $O_p(A_{n-1})$.
The reason to employ $p^{\pm 2}$ here instead of $p^{\pm 1}$
is to avoid $p^{\pm \frac{1}{2}}$ in the forthcoming formulas like
(\ref{btc}) and (\ref{bfi})--(\ref{shnk}) via $q^{\mp 1}$
by (\ref{pa1}).

\begin{remark}\label{re:tl}
Let  $T_{q,n}$ denote the Temperley-Lieb algebra \cite{TL}
generated by 
$t_1, \ldots, t_{n-1}$ obeying the relations
\begin{equation}
\begin{split}
&t_it_j - t_j t_i =0 \quad (|i-j|\ge 2),
\\
&t_i^2= (q+q^{-1})t_i,
\\
&t_it_j t_i = t_i \quad (|i-j|=1).
\end{split}
\end{equation}
Under the relation $p^2=-q^{-2}$ according to (\ref{pa1}), 
it is easy to see that 
\begin{align}\label{btc}
\bt_i \mapsto t_i -\frac{1}{q+q^{-1}}
\end{align}
yields an algebra homomorphism $O_p(A_{n-1}) \rightarrow T_{q,n}$.
The case $p^2=1$ studied in \cite{UI} corresponds to
the singular situation $q+q^{-1}=0$.
\end{remark}

\subsection{Representation $\pi^{\mathrm tr}_l$}

The representation $\pi^{\mathrm tr}_l$ of 
$O_p(A^{(1)}_{n-1})$ on $V_l$ is obtained by the composition 
\begin{align}
\pi^{\mathrm tr}_l:& \; 
O_p(A^{(1)}_{n-1}) \hookrightarrow U_p(A^{(1)}_{n-1})
\rightarrow \mathrm{End} V_l,
\label{ptrk}
\end{align}
where the latter is the $l$-th fundamental representation
(\ref{a:re}) and the former embedding is given by
\begin{align}\label{bfi}
\bt_i &= f_i +p^2k^{-1}_ie_i 
+ \frac{1}{q+q^{-1}}k^{-1}_i
\quad (i \in \Z_n).
\end{align} 
This corresponds to \cite[eq.(34)]{KOY2} with 
$p=-\I\epsilon q^{-1}$ according to (\ref{pa1}).

The summands in (\ref{bfi}) 
are expressed by the local spins (\ref{ls}) as follows:  
\begin{align}
f_i &= 
z^{-\delta_{i,0}}\sigma^+_i\sigma^-_{i+1},
\quad
p^2k^{-1}_ie_i = z^{\delta_{i,0}}\sigma^-_i\sigma^+_{i+1},
\label{ayam}\\
\frac{1}{q+q^{-1}}k^{-1}_i
&= \frac{q+q^{-1}}{4}\sigma^z_i\sigma^z_{i+1}
+\frac{q-q^{-1}}{4}(\sigma^z_i-\sigma^z_{i+1})
-\frac{(q-q^{-1})^2}{4(q+q^{-1})}.
\label{shnk}
\end{align}
The sum of two terms in (\ref{ayam}) with $i=0$ is also written as 
\begin{align}
z^{-1}\sigma^+_n\sigma^-_{1}
+z \sigma^-_n\sigma^+_{1}
= \frac{z+z^{-1}}{4}(\sigma^x_n\sigma^x_1+\sigma^y_n\sigma^y_1)
- \frac{z-z^{-1}}{4\I}(\sigma^x_n\sigma^y_1 - \sigma^y_n\sigma^x_1),
\end{align}
where the second summand is a Dzyaloshinskii-Moriya (DM) interaction term.
The constant term appearing in (\ref{shnk})  
\begin{align}\label{Gam}
\Gamma = -\frac{(q-q^{-1})^2}{4(q+q^{-1})} = 
\frac{(t^2-t^{-2})^2}{4(t^2+t^{-2})}
\end{align}
will be encountered repeatedly in the sequel.

We denote the image of $\bt_i$ by the composition (\ref{ptrk}) 
by $b_i$, i.e., $b_i=\pi^{\mathrm{tr}}_l(\bt_i)$.
Thus it is identified with a local Hamiltonian of XXZ type:
\begin{align}\label{kuwa}
b_i  = z^{-\delta_{i,0}}\sigma^+_i\sigma^-_{i+1}
+z^{\delta_{i,0}}\sigma^-_i\sigma^+_{i+1}
+\frac{q+q^{-1}}{4}\sigma^z_i\sigma^z_{i+1}
+\frac{q-q^{-1}}{4}(\sigma^z_i-\sigma^z_{i+1})
+\Gamma
\quad (i \in \Z_n).
\end{align}

\begin{remark}\label{re:ct}
According to \cite[Prop.2.1]{BB}, setting
\begin{align}\label{bfi2}
\bt_i &= f_i +p^2k^{-1}_ie_i 
+ d_i k^{-1}_i
\quad (i \in \Z_n)
\end{align} 
provides an embedding 
$O_p(A^{(1)}_{n-1}) \hookrightarrow U_p(A^{(1)}_{n-1})$ if and only if 
the coefficients $d_i$ satisfy\footnote{
The condition $a_{ij}=a_{ji}$ is redundant for $A^{(1)}_{n-1}$, but it is 
included for the later use (\ref{bbcon2}) in non simply-laced algebras.}
\begin{align}\label{bbcon1}
d_i\left(d_j^2-\frac{1}{(q+q^{-1})^2}\right)=0 
\quad \forall (i,j)\; \text{ such that }\; (a_{ij}, a_{ji}) = (-1,-1).
\end{align}
The formula (\ref{bfi}) corresponds to
$\forall d_i = \frac{1}{q+q^{-1}}$.
Another choice $\forall d_i = \frac{-1}{q+q^{-1}}$ followed by 
a similarity transformation 
$\bt_i \mapsto \sigma^x \bt_i \sigma^x$ by (\ref{sx})
leads to another representation of $O_p(A_{n-1})$:
\begin{align}\label{kuwa2}
b'_i  = \sigma^+_i\sigma^-_{i+1}
+\sigma^-_i\sigma^+_{i+1}
-\frac{q+q^{-1}}{4}\sigma^z_i\sigma^z_{i+1}
+\frac{q-q^{-1}}{4}(\sigma^z_i-\sigma^z_{i+1})
-\Gamma
\quad (1 \le i < n).
\end{align}
Its constant shift according to (\ref{btc}), i.e., 
\begin{align}
b'_i+\frac{1}{q+q^{-1}}
= \sigma^+_i\sigma^-_{i+1}
+ \sigma^-_i\sigma^+_{i+1}
-\frac{q+q^{-1}}{4}\sigma^z_i\sigma^z_{i+1}
+\frac{q-q^{-1}}{4}(\sigma^z_i-\sigma^z_{i+1})
+\frac{q+q^{-1}}{4}
\end{align}
reproduces the well-known realization  
of the Temperley-Lieb generators 
by an $n$ site spin ${\textstyle \frac{1}{2}}$ chain \cite{PS,NRD,AK}.
\end{remark}

It has been shown \cite{KOY2} that 
the $K$ matrix $K_{\mathrm{tr}}(z)$ (\ref{ktr1})--(\ref{mpktr})
is characterized, up to normalization, by the commutativity 
with the Onsager algebra:
\begin{align}\label{kbtr}
K_{\mathrm{tr}}(z)b_i  = (b_i|_{z\rightarrow z^{-1}})K_{\mathrm{tr}}(z)
\qquad (0 \le i < n),
\end{align}
where the replacement $z\rightarrow z^{-1}$ matters only for $i=0$.

Set
\begin{equation}\label{ha}
H_{\mathrm{tr}}(z) = b_0 + b_1 + \cdots + b_{n-1}
=\sum_{i \in \Z_n} 
\Bigl(
z^{-\delta_{i,0}}\sigma^+_i\sigma^-_{i+1}
+z^{\delta_{i,0}}\sigma^-_i\sigma^+_{i+1}
+\frac{q+q^{-1}}{4}\sigma^z_i\sigma^z_{i+1}\Bigr)
+ n\Gamma,
\end{equation}
where the $z$-dependence comes only from $b_0$.
We have taken the coefficients of $b_i$'s so that 
the sum eliminates the $\sigma^z_i$-linear terms in (\ref{kuwa}), and therefore
$\sigma^x H_{\mathrm{tr}}(z)\sigma^x = H_{\mathrm{tr}}(z^{-1})$
holds with $\sigma^x$ defined by (\ref{sx}).
Then (\ref{kbtr}) and (\ref{kve}) lead to the commutativity 
\begin{align}\label{kvh:A}
[K^\vee_{\mathrm{tr}}(z), H_{\mathrm{tr}}(z)] = 0.
\end{align}
To construct higher order commuting Hamiltonians 
within the Onsager algebra $O_p(A^{(1)}_{n-1})$
is an outstanding problem whose solution has been 
known only at $p=1$ \cite{UI,BCP}.
See also the ending remarks in Section \ref{sec:sum}.
As far as $O_p(A^{(1)}_{n-1})$ is concerned, it may be useful  
to combine Remark \ref{re:tl} and \cite{HMR}.

Let us comment on the hermiticity of the Hamiltonians.
The local ones $b_0,\ldots, b_{n-1}$ (\ref{kuwa})
are all hermite if and only if $|z|=1$ and $q \in \R$.
When  $|z|=1$ and $q \in \I \R$, 
they are hermite except for the summand 
$\frac{q-q^{-1}}{4}(\sigma^z_i-\sigma^z_{i+1})$
representing a pure imaginary magnetic field.
On the other hand, 
$H_{\mathrm{tr}}(z)$ (\ref{ha}) is hermite 
if and only if $|z|=1$ and either $q \in \R$ or $q \in \I  \R$.
A similar feature holds for $\mathfrak{g}$ other than $A^{(1)}_{n-1}$. 

\begin{remark}\label{re:aoda}
It is possible to formulate an $n$-parameter version of the above result.
This is due to the algebra automorphism 
$e_i \mapsto z_i e_i$,
$f_i \mapsto z_i^{-1} f_i$,  
$k_i \mapsto k_i$
involving the nonzero parameters $z_i\,(i \in \Z_n)$.
Alternatively, one may keep (\ref{bfi}) 
and modify the representation (\ref{a:re}) into
\begin{align}
e_j|\alb \rangle 
= z_j|\alb-{\bf e}_j+{\bf e}_{j+1}\rangle, 
\quad
f_j|\alb \rangle 
= z_j^{-1}|\alb+{\bf e}_j-{\bf e}_{j+1}\rangle,
\quad
k_j |\alb \rangle
= p^{2(\alpha_{j+1}-\alpha_{j})}|\alb\rangle
\quad (j \in \Z_n).
\end{align}
Then (\ref{ha}) is changed into
\begin{equation}\label{han}
H_{\mathrm{tr}}(z_0,\ldots, z_{n-1}) = b_0 + \cdots + b_{n-1}
=\sum_{i \in \Z_n} 
\Bigl(
z_i^{-1}\sigma^+_i\sigma^-_{i+1}
+z_i\sigma^-_i\sigma^+_{i+1}
+\frac{q+q^{-1}}{4}\sigma^z_i\sigma^z_{i+1}\Bigr)
+ n\Gamma.
\end{equation}
The choice $z_i = z$ for all $i\in \Z_n$ is
the model involving uniform DM terms studied in \cite{AW}.
Introduce 
$K^\vee_{\mathrm{tr}}(z_0,\ldots, z_{n-1})= \sigma^x 
K_{\mathrm{tr}}(z_0,\ldots, z_{n-1})$ similarly to (\ref{kve}),
where elements of the latter is defined by generalizing (\ref{mpktr}) to
\begin{align}
K_{\mathrm{tr}}(z_0,z_1,\ldots, z_{n-1})^{\beb}_{\alb}
= \mathrm{Tr}(z_0^{\bf h}K^{\beta_1}_{\alpha_1}z_1^{\bf h}K^{\beta_2}_{\alpha_2}
\cdots z_{n-1}^{\bf h}K^{\beta_n}_{\alpha_n})
\end{align}
up to overall normalization.
Then the following commutativity is valid:
\begin{align}
[K^\vee_{\mathrm{tr}}(z_0,\ldots, z_{n-1}), H_{\mathrm{tr}}(z_0,\ldots, z_{n-1})]=0.
\end{align}
This kind of multi-parameter generalizations are possible also for   
$U_p$ treated in later sections, although they will be omitted  
for simplicity. 
\end{remark}

\section{Spectral decomposition of $K_{\mathrm{tr}}$ by $O_p(A_{n-1})$}
\label{sec:spda}

The classical part $O_p(A_{n-1})$ 
of the Onsager algebra $O_p(A^{(1)}_{n-1})$ 
introduced in Section \ref{ss:oaa1}
has the representation 
\begin{align}
O_p(A_{n-1}) \hookrightarrow O_p(A^{(1)}_{n-1})
\overset{\pi^{\mathrm{ tr}}_l}{\longrightarrow}
\mathrm{End} V_l\quad (0 \le l \le n).
\end{align}
We denote this restriction also by $\pi^{\mathrm{ tr}}_l$.
The relation (\ref{kbtr}) with $i \neq 0$ tells the commutativity
\begin{align}
[K_{\mathrm{tr}}(z), \pi^{\mathrm{ tr}}_l(O_p(A_{n-1}))]=0.
\end{align}
The representation $\pi^{\mathrm{ tr}}_l$ of $O_p(A^{(1)}_{n-1})$ on $V_l$ is 
irreducible \cite{KOY2}.
On the other hand it is not so with respect to the 
classical subalgebra $O_p(A_{n-1})$.
The $K$ matrix $K_{\mathrm{tr}}(z)$ 
should be a scalar on each irreducible component.
For instance when $(n,l)=(4,2)$,
it acts on the $\binom{4}{2}=6$ dimensional space $V_2$, and 
its eigenvalues read
\begin{align}
1,  \;\;\;  1, \;\;\;  \frac{q^2-z}{-1+q^2z}, \;\;\; \frac{q^2-z}{-1+q^2z}, 
\;\;\; \frac{q^2-z}{-1+q^2z}, \;\;\; 
\frac{(q^2-z)(q^4-z)}{(1-q^2z)(1-q^4z)}.
\end{align}
The multiplicities $2, 3$ and  $1$ here are equal to the 
Kostka numbers 
$\binom{4}{2}-\binom{4}{1},
\binom{4}{1}-\binom{4}{0}$ and 
$\binom{4}{0}$, respectively.
Systematizing such investigations leads to 
the conjecture that there are irreducible $O_p(A_{n-1})$ modules
$W_{l,j}$ with $0 \le l \le {\textstyle \frac{n}{2}}$ or 
${\textstyle \frac{n}{2}}< l  \le n$ 
having the properties (i), (ii) and (iii) described below:

\vspace{0.2cm}
(i)  $V_0, V_1, \ldots, V_n$ are decomposed as
\begin{align}
V_l &= \begin{cases}
W_{l,l} \oplus W_{l,l-1} \oplus \cdots \oplus W_{l,0},
\quad &(0 \le l \le {\textstyle \frac{n}{2}}),
\\ 
W_{l, l} \oplus W_{l,l+1} \oplus \cdots \oplus W_{l,n}
\quad &({\textstyle \frac{n}{2}}< l  \le n),
\end{cases}
\label{afk}\\
\dim W_{l,j} & = \begin{cases}
\binom{n}{j}-\binom{n}{j-1}\quad
&(0 \le j \le l \le {\textstyle \frac{n}{2}}),
\\
\binom{n}{j}-\binom{n}{j+1}\quad
&({\textstyle \frac{n}{2}}<l \le j \le n).
\end{cases}
\end{align}
This is consistent with $\dim V_l = \binom{n}{l}$ and 
satisfies $\dim W_{l,j} = \dim W_{n-l, n-j}$.
For convenience when $n$ is even, 
we also define $W_{\frac{n}{2},j}$ with $j>{\textstyle \frac{n}{2}}$
by setting $W_{\frac{n}{2},j} = W_{\frac{n}{2},n-j}$ for all $0 \le j \le n$.

(ii) The decomposition $K_{\mathrm{tr}}(z) 
\in \bigoplus_{l=0}^n \mathrm{Hom}(V_l,  V_{n-l})$ in (\ref{ktrd}) 
is refined into
\begin{align}
K_{\mathrm{tr}}(z) &\in \bigoplus_{0 \le j \le l \le \frac{n}{2}}
\mathrm{Hom}(W_{l,j}, W_{n-l,n-j})
\oplus 
\bigoplus_{\frac{n}{2} < l \le j \le n}
\mathrm{Hom}(W_{l,j}, W_{n-l,n-j}),
\end{align}
where each component is an isomorphism of 
$O_p(A_{n-1})$ modules.
 
(iii) There exists a basis $\{\xi^{l,j}_i \mid 1 \le i \le \dim W_{l,j}\}$
of $W_{l,j}$ in terms of which the isomorphism in (ii) 
is explicitly described as the spectral decomposition:
\begin{align}
K_{\mathrm{tr}}(z)
&= \bigoplus_{0 \le j \le l \le \frac{n}{2}} \rho_{l,j}(z)P_{l,j}
\;\;\oplus 
\bigoplus_{\frac{n}{2} < l \le j \le n} \rho_{l,j}(z)P_{l,j},
\\
P_{l,j}\, \xi^{l',j'}_i  &= \delta_{l,l'}\delta_{j,j'}\xi^{n-l,n-j}_i,
\\
\rho_{l,j}(z) &=(-z)^{|l-j|}
\frac{(q^{|n-2l|+2}z^{-1};q^2)_\infty (q^{|n-2j|+2}z;q^2)_\infty}
{(q^{|n-2l|+2}z;q^2)_\infty (q^{|n-2j|+2}z^{-1};q^2)_\infty}.
\label{cdko}
\end{align}
We have used infinite products in order to make the formula uniform.
However all the eigenvalues of the $K$ matrices appearing here and 
in what follows are rational functions of $z$.
It is easy to see $\rho_{l,j}(z)\rho_{n-l,n-j}(z^{-1})=1$,
which is consistent with (\ref{petb}).

In view of Remark \ref{re:tl}, we expect that 
$W_{l,j}\, (0 \le j \le l \le \frac{n}{2})$ 
is the irreducible representation $V_{[n-j,j]}$ of the Temperley-Lieb algebra
$T_{q,n}$ labeled with the two row Young diagram 
$[n-j,j]$ in \cite[p126]{GHJ}.
The decomposition (\ref{afk}) corresponds to 
\cite[eq.(57)]{AK}.

\section{Types other than $U_p(A^{(1)}_{n-1})$: general features}
\label{sec:tatsuki}

This brief section is a guide to Sections \ref{sec:D2}--\ref{sec:spdd}
where contents analogous to $U_p(A^{(1)}_{n-1})$ case 
in Section \ref{sec:A}--\ref{sec:spda}
will be presented individually for the other $U_p$ under consideration.
They consist of so many cases that one may wonder if it is 
possible to grasp them in a unified manner.
Our aim here is to indicate how to do so at least partially.
We note that these variety of cases have 
originated in the solutions of the reflection equation 
listed in \cite[Sec.6]{KP}
and the corresponding coideals in \cite[App.B]{KOY2}.  

For convenience we set 
\begin{align}
\mathfrak{g}^{1,1} = D^{(2)}_{n+1},\quad
\mathfrak{g}^{2,1} = B^{(1)}_{n},\quad
\mathfrak{g}^{1,2} = \tilde{B}^{(1)}_{n},\quad
\mathfrak{g}^{2,2} = D^{(1)}_n.
\end{align}
The superscript $r$ in $\mathfrak{g}^{r,r'}$
indicates that the Dynkin diagram around the $0$ th vertex is 
an outward double arrow for $r=1$ and trivalent for $r=2$.
The shape around the $n$ th vertex is specified by $r'$ similarly.
The quantity $p_i$ defined after (\ref{uqdef}) is 
written as $p_0=p^r$, $p_i = p^2\,(0<i<n)$ and $p_n = p^{r'}$.

For each $\mathfrak{g}^{r,r'}$, 
we will consider the quantum affine algebra 
$U_p(\mathfrak{g}^{r,r'})$ and the Onsager algebra 
$O_p(\mathfrak{g}^{r,r'})$ \cite{BB}.
The Serre relations in $U_p(\mathfrak{g}^{r,r'})$ read
\begin{align}
&e_ie_j - e_je_i = 0 \quad (a_{ij}=0),
\label{sr1}\\
&e_i^2e_j -(p^2+p^{-2})e_ie_je_i+e_je_i^2 = 0\quad (a_{ij}=-1),
\label{sr2}\\
&e_i^3e_j - (p^2+1+p^{-2})e_i^2e_je_i + (p^2+1+p^{-2})e_ie_je^2_i-e_j e_i^3
= 0 \quad (a_{ij}=-2)
\label{sr3}
\end{align}
and the same ones for $f_j$'s.
The other relations have been already given in (\ref{uqdef}).

The Onsager algebra 
$O_p(\mathfrak{g}^{r,r'})$ is generated by $\bt_0,\ldots, \bt_n$
obeying modified $p$-Serre relations \cite{BB}:
\begin{align}
&\bt_i\bt_j - \bt_j\bt_i = 0 \quad (a_{ij}=0),
\label{bbre1}\\
&\bt_i^2\bt_j -(p^2+p^{-2})\bt_i\bt_j\bt_i+\bt_j\bt_i^2 = \bt_j \quad 
(a_{ij}=-1),
\label{bbre2}\\
&\bt_i^3\bt_j - (p^2+1+p^{-2})\bt_i^2\bt_j\bt_i 
+ (p^2+1+p^{-2})\bt_i\bt_j \bt^2_i-\bt_j \bt_i^3
=(p+p^{-1})^2(\bt_i\bt_j-\bt_j\bt_i) \quad (a_{ij}=-2).
\label{bbre3}
\end{align}
Except for (\ref{sr3}) and (\ref{bbre3}) which are void for 
the simply-laced 
$U_p(\mathfrak{g}^{2,2})$ and $O_p(\mathfrak{g}^{2,2})$,
these relations are formally the same with those in type $A^{(1)}_{n-1}$.
In terms of commutators $[X,Y]=[X,Y]_1$, $[X,Y]_r= XY-rYX$, 
the relations (\ref{bbre1})--(\ref{bbre3}) are written more compactly as
\begin{align}
&[\bt_i, \bt_j]=0\quad (a_{ij}=0),
\\
& [\bt_i, [\bt_i, \bt_j]_{p^2}]_{p^{-2}} = \bt_j\quad (a_{ij}=-1),
\\ 
&[\bt_i, [\bt_i, [\bt_i, \bt_j]_{p^2}]_{p^{-2}}]=(p+p^{-1})^2
[\bt_i,\bt_j]\quad (a_{ij}=-2).
\label{kcd}
\end{align}
The quartic relation of the form (\ref{kcd}) with $p^2=1$ 
is often referred to as the Dolan-Grady condition \cite{DG}.
It is typical for the situation  $a_{ij}=-2$, which was
indeed utilized to reformulate the original Onsager algebra for $A^{(1)}_1$
\cite{On} by only two generators.
The Onsager algebra $O_p(D^{(1)}_n)$ with $p=1$ was
introduced in \cite{DU}.
It is an interesting open question if there is an analogue of Remark \ref{re:tl} 
for $\mathfrak{g} \neq A^{(1)}_{n-1}$ related to 
a boundary extension of the Temperley-Lieb algebra like \cite{DN}.

We will deal with the representations of $O_p(\mathfrak{g}^{r,r'})$
constructed as
\begin{align}\label{scm}
\pi^{r,r'}_{k,k'}:\; 
O_p(\mathfrak{g}^{r,r'}) \hookrightarrow 
U_p(\mathfrak{g}^{r,r'}) \rightarrow \mathrm{End} V
\qquad ((r,k), (r',k')=(1,1), (1,2), (2,2))
\end{align}
with $V = (\C^2)^{\otimes n}$.
Thus there are {\em nine} cases to consider.
We remark that the strange condition $r \le k, r' \le k'$ originates in 
(\ref{bv}) to validate Theorem \ref{th:bv}, which was a key in the 
3D approach \cite{KP}.
The latter arrow in (\ref{scm}) is the spin representations of 
$U_p(\mathfrak{g}^{r,r'})$ which will be specified in later sections.
They carry a spectral parameter $z$.
The former embedding depends on $k,k'$ and is given by
\begin{align}
\bt_0 &= f_0 + p^r k_0^{-1}e_0 + d^{r}_{k}k_0^{-1},
\label{bg0}\\
\bt_i &= f_i + p^2 k_i^{-1}e_i + \frac{1}{q+q^{-1}}k^{-1}_i\;\;
(0 < i < n),
\\
\bt_n &= f_n + p^{r'}k_n^{-1}e_n + d^{r'}_{k'}k_n^{-1},
\\
d^1_1 & = \epsilon\frac{q^{\frac{1}{2}}+q^{-\frac{1}{2}}}{q+q^{-1}},
\quad
d^1_2 = 0,\quad
d^2_2 = \frac{1}{q+q^{-1}}.
\label{drk}
\end{align}
See (\ref{pa1}) for the relations among the parameters $p, q, \epsilon$ etc.
Recall also that $p_0,\ldots, p_n$ were specified after (\ref{uqdef}).
In general, according to \cite[Prop.2.1]{BB}, setting 
$\bt_i = 
f_i + p_i k_i^{-1}e_i + d_ik^{-1}_i\,(0\le i \le n)$
provides an embedding 
$O_p(\mathfrak{g}^{r,r'}) \hookrightarrow U_p(\mathfrak{g}^{r,r'})$ if and only if 
the following condition is satisfied:
\begin{align}\label{bbcon2}
(\ref{bbcon1}) \;\;\text{ and }\;\; 
d_i\left(d_j^2-\frac{1}{(q+q^{-1})^2}\right)=0 
\quad \forall (i,j)\; \text{ such that }\; (a_{ij},a_{ji}) = (-2,-1).
\end{align}
One can check that (\ref{bg0})--(\ref{drk}) fulfills this and 
the relations (\ref{bbre1})--(\ref{bbre3}) directly.

As in $U_p(A^{(1)}_{n-1})$, 
we shall write $b_i= \pi^{r,r'}_{k,k'}(\bt_i)$ to mean the representation 
(\ref{scm}) of $\bt_i \in O_p(\mathfrak{g}^{r,r'})$.
Its dependence on $(k,k')$ should not be forgotten although it is 
suppressed in the notation for simplicity.
Among $b_0, b_1, \ldots, b_n$,  there is a special (affine) one 
$b_0$ which includes the spectral parameter $z$
built in the spin representations.

It has been shown \cite{KOY2} 
that the $K$ matrix $\tilde{K}_{k,k'}(z)$ 
is characterized, up to normalization, by the commutativity with the Onsager algebra:
\begin{align}\label{kkoy}
\tilde{K}_{k,k'}(z)b_i  
= (b_i|_{z\rightarrow z^{-1}})\tilde{K}_{k,k'}(z)
\qquad (0 \le i \le n),
\end{align} 
where $\tilde{K}_{k,k'}(z)$ has been defined in (\ref{ktd}).

It turns out that the analogue of $H_{\mathrm{tr}}(z)$ in (\ref{ha}) 
can be constructed for 
the representation $\pi^{r,r'}_{k,k'}$ if and only if $(r,r')=(k,k')$.
In fact, for the generators $b_0,\ldots, b_n$ 
in $\pi^{k,k'}_{k,k'}(O_p(\mathfrak{g}^{k,k'}))$, 
it is possible to choose the constant ($z$-independent) coefficients 
$\kappa_0,\ldots, \kappa_n$ so that
\begin{align}
H_{k,k'}(z) = \kappa_0 b_0 + \kappa_1 b_1 + \cdots + \kappa_n b_n
\end{align}
becomes free from $\sigma_i^z$-linear terms and fulfills 
$\sigma^x H_{k,k'}(z) \sigma^x = H_{k,k'}(z^{-1})$.
As a result, (\ref{kkoy}) leads to 
\begin{align}
[K^\vee_{k,k'}(z), H_{k,k'}(z)]=0,
\end{align} 
where $K^\vee_{k,k'}(z)$ has been introduced in (\ref{kve}).
The concrete forms of $H_{k,k'}(z)$ will be presented 
in (\ref{h11}), (\ref{h21}), (\ref{h12}) and (\ref{h22}).

The local Hamiltonians $b_0, \ldots, b_n$ are all hermite 
if and only if $|z|=1$ and $t^{\min(k,k')} \in \R$,
where $t$ is related to $p$ and $q$ as in (\ref{pa1}).
When $|z|=1$ and $t^{\min(k,k')} \in \I \R$,
some of them acquire a pure imaginary magnetic field term.
The Hamiltonian $H_{k,k'}(z)$ is hermite if and only if
$|z|=1$ and either $t^{\min(k,k')} \in \R$ or $t^{\min(k,k')} \in \I \R$. 

Next let us motivate Section \ref{sec:spdb} and \ref{sec:spdd}.
In view of the Dynkin diagrams, 
it is natural to denote the subalgebra  
of $O_p(\mathfrak{g}^{r,r'})$ generated by $\bt_1, \ldots, \bt_n$
by $O_p(B_n)$ for $r'=1$ and $O_p(D_n)$ for $r'=2$.
By inspection it is easy to see that
$\pi^{r,1}_{k,1}$ of $O_p(\mathfrak{g}^{r,1})$ defines the same
representation of $O_p(B_n)$ for any choice $(r,k)=(1,1), (1,2), (2,2)$.
This common representation will naturally be denoted by $\pi^1_1$.
The relation (\ref{kkoy}) implies that 
$\tilde{K}_{k,1}(z)$ commutes with $O_p(B_n)$ 
in the representation $\pi^1_1$.
The spectral decomposition of $K_{k,1}(z)$ with respect to 
$\pi^1_1(O_p(B_n))$ will be given in Section \ref{sec:spdb}.
Similarly, $\pi^{r,2}_{k,2}$ of $O_p(\mathfrak{g}^{r,2})$
yields the same representation of $O_p(D_n)$
for $(r,k)=(1,1), (1,2), (2,2)$.
It will be denoted by $\pi^2_2$.
The relation (\ref{kkoy}) also implies that 
$\tilde{K}_{k,2}(z)$ commutes with $O_p(D_n)$ 
in the representation $\pi^2_2$.
The spectral decomposition of $K_{k,2}(z)$ with respect to 
$\pi^2_2(O_p(D_n))$ will be given in Section \ref{sec:spdd}.
(We do not treat $r'\neq k'$ case to avoid technical complexity.)
 
\section{$O_p(D^{(2)}_{n+2})$ Hamiltonians}\label{sec:D2}

\subsection{$U_p(D^{(2)}_{n+1})$ and spin representation}

The Dynkin diagram and the Cartan matrix are given by
\begin{align*}
\begin{picture}(200,50)
\put(50,0){
\put(30,24){$\mathfrak{g}^{1,1}= D^{(2)}_{n+1}$}
\multiput( 0,0)(20,0){2}{\circle{6}}
\multiput(80,0)(20,0){2}{\circle{6}}
\put(23,0){\line(1,0){14}}
\put(62.5,0){\line(1,0){14}}
\multiput(2.85,-1)(0,2){2}{\line(1,0){14.3}} 
\multiput(82.85,-1)(0,2){2}{\line(1,0){14.3}} 
\multiput(39,0)(4,0){6}{\line(1,0){2}} 
\put(10,0.2){\makebox(0,0){$<$}}
\put(90,0.2){\makebox(0,0){$>$}}
\put(0,-6){\makebox(0,0)[t]{$0$}}
\put(20,-6){\makebox(0,0)[t]{$1$}}
\put(80,-6){\makebox(0,0)[t]{$n\!\! -\!\! 1$}}
\put(100,-7.8){\makebox(0,0)[t]{$n$}}
}
\end{picture}
(a_{ij})_{0\le i,j \le n}= 
\begin{pmatrix}
2 & -2  & \cdots &  0 & 0
\\
-1 & 2  & \cdots  & & 0 
\\
\vdots   & \ddots & \ddots & \ddots &  \vdots \\
0 &  &  & 2 &  -1
\\
0 & 0 & \cdots  & -2 & 2
\end{pmatrix}.
\end{align*}

The spin representation on $V$ is given by 
\begin{alignat}{3}
e_0|\alb \rangle 
&= z|\alb+{\bf e}_1\rangle,
&\quad
f_0|\alb \rangle 
&= z^{-1}|\alb-{\bf e}_1\rangle,
&\quad
k_0|\alb \rangle 
&= p^{2\alpha_1-1}|\alb\rangle,
\label{d2:re1}\\
e_j|\alb \rangle 
&= |\alb-{\bf e}_j+{\bf e}_{j+1}\rangle, 
&\quad
f_j|\alb \rangle 
&= |\alb+{\bf e}_j-{\bf e}_{j+1}\rangle,
&\quad
k_j |\alb \rangle
&= p^{2(\alpha_{j+1}-\alpha_{j})}|\alb\rangle\quad(0<j<n),
\\
e_n|\alb \rangle 
&= |\alb-{\bf e}_n\rangle,
&\quad
f_n|\alb \rangle 
&= |\alb+{\bf e}_n\rangle,
&\quad
k_n|\alb \rangle 
&= p^{1-2\alpha_n}|\alb\rangle.
\label{d2:re2}
\end{alignat}

\subsection{Onsager algebra $O_p(D^{(2)}_{n+1})$
and the classical part $O_p(B_n)$}\label{ss:oad2}
The algebra $O_p(D^{(2)}_{n+1})$ is generated by $\bt_0,\ldots, \bt_n$
obeying (\ref{bbre1})--(\ref{bbre3}).
The classical part of $D^{(1)}_{n+1}$ without the vertex 0 is $B_n$.
Thus the subalgebra of $O_p(D^{(2)}_{n+1})$ generated by 
$\bt_1,\ldots, \bt_n$ is the Onsager algebra for $B_n$.
We denote it by $O_p(B_n)$.

\subsection{Representations $\pi^{1,1}_{1,1}$}

The representation $\pi^{1,1}_{1,1}$ of $O_p(D^{(2)}_{n+1})$ 
on $V$ in (\ref{Vdef}) 
is obtained by the composition 
\begin{align}\label{p1111}
\pi^{1,1}_{1,1}:
O_p(D^{(2)}_{n+1}) \hookrightarrow U_p(D^{(2)}_{n+1})  
\rightarrow \mathrm{End} V,
\end{align}
where the latter is the spin representation (\ref{d2:re1})-(\ref{d2:re2}) 
and the former embedding (\ref{bg0})--(\ref{drk}) reads as
\begin{align}
\bt_0 
&= f_0 + p k_0^{-1}e_0
-\I \epsilon\mu\frac{t-t^{-1}}{t^2+t^{-2}}k^{-1}_0,
\label{re1}\\
\bt_i &= f_i +p^{2}k^{-1}_ie_i 
- \frac{1}{t^2+t^{-2}}k^{-1}_i
\quad (0 < i <n),
\\
\bt_n 
&= f_n + p^{} k^{-1}_ne_n 
-\I \epsilon\mu\frac{t-t^{-1}}{t^2+t^{-2}}k^{-1}_n.
\label{re3}
\end{align}
This corresponds to  \cite[eqs.(167)-(169)]{KOY2} with 
$s=s' = q^\hf$ according to \cite[eq.(96)]{KOY2}.
These generators are represented as local Hamiltonians:
\begin{align}
b_0&= z \sigma^+_1+z^{-1}\sigma^-_1-\mu\frac{t-t^{-1}}{2}\sigma^z_1
- \mu\frac{(t-t^{-1})(t^2-t^{-2})}{2(t^2+t^{-2})},
\label{b11110}\\
b_i & = \sigma^+_i\sigma^-_{i+1}+\sigma^-_i\sigma^+_{i+1}
-\frac{t^2+t^{-2}}{4}\sigma^z_i\sigma^z_{i+1}-
\frac{t^2-t^{-2}}{4}(\sigma^z_i-\sigma^z_{i+1})
+\frac{(t^2-t^{-2})^2}{4(t^2+t^{-2})}\;\;(0<i<n),
\\
b_n &= \sigma^x_n + \mu\frac{t-t^{-1}}{2}\sigma^z_n-
\mu\frac{(t-t^{-1})(t^2-t^{-2})}{2(t^2+t^{-2})}.
\label{b1111n}
\end{align}
They commute with the $K$ matrix (\ref{ktd}) up to $z$ as
\cite{KOY2}:
\begin{align}\label{kb1111}
\tilde{K}_{1,1}(z)b_i  = (b_i|_{z\rightarrow z^{-1}})\tilde{K}_{1,1}(z)
\qquad (0 \le i \le n).
\end{align} 

Set
\begin{equation}\label{h11}
\begin{split}
H_{1,1}(z) &= -\frac{\mu(t+t^{-1})}{2} b_0 + b_1 + \cdots + b_{n-1}
-\frac{\mu(t+t^{-1})}{2}b_n
\\
&= -\frac{\mu(t+t^{-1})}{2}(z\sigma^+_1+z^{-1}\sigma^-_1+\sigma^x_n)
+ \sum_{i=1}^{n-1}\Bigl(
\sigma^+_i\sigma^-_{i+1}+ \sigma^-_i\sigma^+_{i+1}
- \frac{t^2+t^{-2}}{4}\sigma^z_i\sigma^z_{i+1}
\Bigr) + (n+1)\Gamma.
\end{split}
\end{equation}
It satisfies $\sigma^x H_{1,1}(z) \sigma^x = H_{1,1}(z^{-1})$, therefore 
(\ref{kb1111}) and (\ref{kve}) lead to the commutativity
\begin{align}
[K^\vee_{1,1}(z), H_{1,1}(z)]=0.
\end{align}
The Hamiltonian $H_{1,1}(z)$ has appeared for example in 
\cite[eq.(1.3)]{N1} with 
$\beta_\pm=\theta_-=0$, 
$\mathrm{e}^{\theta_+} = z, 
\mathrm{e}^{\eta} = -t^2$,
$\sinh \alpha_\pm = \frac{\mu(t-t^{-1})}{2}$.

\subsection{Representation $\pi^{1,1}_{2,1}$}

The representation $\pi^{1,1}_{2,1}$ of $O_p(D^{(2)}_{n+1})$ 
on $V$ in (\ref{Vdef}) 
is obtained by the composition 
\begin{align}\label{p1121}
\pi^{1,1}_{2,1}:
O_p(D^{(2)}_{n+1}) \hookrightarrow U_p(D^{(2)}_{n+1})  
\rightarrow \mathrm{End} V,
\end{align}
where the latter is the spin representation (\ref{d2:re1})-(\ref{d2:re2}) 
and the former embedding (\ref{bg0})--(\ref{drk}) reads as
\begin{align}
\bt_0 
&= f_0 + p k_0^{-1}e_0
\label{re7}\\
\bt_i &= f_i +p^{2}k^{-1}_ie_i 
-\frac{1}{t^2+t^{-2}}k^{-1}_i
\quad (0 < i <n),
\\
\bt_n&= f_n + p^{} k^{-1}_ne_n 
-\I\epsilon\mu\frac{t-t^{-1}}{t^2+t^{-2}}k^{-1}_n.
\label{re9}
\end{align} 
This corresponds to \cite[eqs.(167), (172)-(173)]{KOY2}
with $s=s'= q^\hf$.
These generators are represented as local Hamiltonians:
\begin{align}
b_0&= z \sigma^+_1+z^{-1}\sigma^-_1,
\\
b_i & = \sigma^+_i\sigma^-_{i+1}+\sigma^-_i\sigma^+_{i+1}
-\frac{t^2+t^{-2}}{4}\sigma^z_i\sigma^z_{i+1}-
\frac{t^2-t^{-2}}{4}(\sigma^z_i-\sigma^z_{i+1})
+\frac{(t^2-t^{-2})^2}{4(t^2+t^{-2})}\;\;\;(0<i<n),
\\
b_n &= \sigma^x_n + \mu\frac{t-t^{-1}}{2}\sigma^z_n-
\mu\frac{(t-t^{-1})(t^2-t^{-2})}{2(t^2+t^{-2})}.
\end{align}
They commute with the $K$ matrix (\ref{ktd}) up to $z$ as
\cite{KOY2}:
\begin{align}\label{kb1121}
\tilde{K}_{2,1}(z) b_i  = (b_i|_{z\rightarrow z^{-1}})\tilde{K}_{2,1}(z)
\qquad (0 \le i \le n).
\end{align} 

\subsection{Representation $\pi^{1,1}_{1,2}$}

The representation $\pi^{1,1}_{1,2}$ of $O_p(D^{(2)}_{n+1})$ 
on $V$ in (\ref{Vdef}) 
is obtained by the composition 
\begin{align}\label{p1112}
\pi^{1,1}_{1,2}:
O_p(D^{(2)}_{n+1}) \hookrightarrow U_p(D^{(2)}_{n+1})  
\rightarrow \mathrm{End} V,
\end{align}
where the latter is the spin representation (\ref{d2:re1})-(\ref{d2:re2}) 
and the former embedding (\ref{bg0})--(\ref{drk}) reads as
\begin{align}
\bt_0 
&= f_0 + p k_0^{-1}e_0
-\I \epsilon\mu\frac{t-t^{-1}}{t^2+t^{-2}}k^{-1}_0,
\label{re10}\\
\bt_i &= f_i +p^2k^{-1}_ie_i 
- \frac{1}{t^2+t^{-2}}k^{-1}_i
\quad (0 < i <n),
\\
\bt_n 
&= f_n + p^{} k^{-1}_ne_n.
\label{re12}
\end{align}
This corresponds to  \cite[eqs.(167), (176)-(177)]{KOY2} with 
$s=s' = q^\hf$ according to \cite[eq.(96)]{KOY2}.
These generators are represented as local Hamiltonians:
\begin{align}
b_0&= z \sigma^+_1+z^{-1}\sigma^-_1-\mu\frac{t-t^{-1}}{2}\sigma^z_1
- \mu\frac{(t-t^{-1})(t^2-t^{-2})}{2(t^2+t^{-2})},
\\
b_i & = \sigma^+_i\sigma^-_{i+1}+\sigma^-_i\sigma^+_{i+1}
-\frac{t^2+t^{-2}}{4}\sigma^z_i\sigma^z_{i+1}-
\frac{t^2-t^{-2}}{4}(\sigma^z_i-\sigma^z_{i+1})
+\frac{(t^2-t^{-2})^2}{4(t^2+t^{-2})}\;\;(0<i<n),
\\
b_n &= \sigma^x_n.
\end{align}
They commute with the $K$ matrix (\ref{ktd}) up to $z$ as
\cite{KOY2}:
\begin{align}\label{kb1112}
\tilde{K}_{1,2}(z) b_i  = (b_i|_{z\rightarrow z^{-1}})\tilde{K}_{1,2}(z)
\qquad (0 \le i \le n).
\end{align} 

\subsection{Representation $\pi^{1,1}_{2,2}$}

The representation $\pi^{1,1}_{2,2}$ of $O_p(D^{(2)}_{n+1})$ 
on $V$ in (\ref{Vdef}) is obtained by the composition 
\begin{align}\label{p1122}
\pi^{1,1}_{2,2}:
O_p(D^{(2)}_{n+1}) \hookrightarrow U_p(D^{(2)}_{n+1})  
\rightarrow \mathrm{End} V,
\end{align}
where the latter is the spin representation (\ref{d2:re1})-(\ref{d2:re2}) 
and the former embedding (\ref{bg0})--(\ref{drk}) reads as 
\begin{align}
\bt_0 
&= f_0 + p k_0^{-1}e_0
\label{re4}\\
\bt_i &= f_i +p^{2}k^{-1}_ie_i 
+ \frac{1}{q+q^{-1}}k^{-1}_i
\quad (0 < i <n),
\\
\bt_n 
&= f_n + p^{} k^{-1}_ne_n.
\label{re6}
\end{align} 
This corresponds to \cite[eqs.(167), (184)-(185)]{KOY2}
with $s=s'= q^\hf$.
These generators are represented as local Hamiltonians:
\begin{align}
b_0&= z \sigma^+_1+z^{-1}\sigma^-_1,
\label{b11220}\\
b_i & = \sigma^+_i\sigma^-_{i+1}+\sigma^-_i\sigma^+_{i+1}
+\frac{q+q^{-1}}{4}\sigma^z_i\sigma^z_{i+1}
+\frac{q-q^{-1}}{4}(\sigma^z_i-\sigma^z_{i+1})
-\frac{(q-q^{-1})^2}{4(q+q^{-1})}\;\;\;(0<i<n),
\\
b_n &= \sigma^x_n.
\label{b1122n}
\end{align}
They commute with the $K$ matrix (\ref{ktd}) up to $z$ as
\cite{KOY2}:
\begin{align}\label{kb1122}
\tilde{K}_{2,2}(z) b_i  = (b_i|_{z\rightarrow z^{-1}})\tilde{K}_{2,2}(z)
\qquad (0 \le i \le n).
\end{align}

\section{$O_p(B^{(1)}_n)$ Hamiltonians}\label{sec:B}

\subsection{$U_p(B^{(1)}_n)$ and spin representations}
The Dynkin diagram and the Cartan matrix are given by
\begin{align*}
\begin{picture}(200,50)
\put(50,0){
\put(35,24){$\mathfrak{g}^{2,1}=B^{(1)}_{n}$}
\put(6,14){\circle{6}}\put(6,-14){\circle{6}}
\put(20,0){\circle{6}}
\multiput(80,0)(20,0){2}{\circle{6}}
\put(23,0){\line(1,0){14}}
\put(62.5,0){\line(1,0){14}}
\put(18,3){\line(-1,1){9}} \put(18,-3){\line(-1,-1){9}}
\multiput(82.85,-1)(0,2){2}{\line(1,0){14.3}} 
\multiput(39,0)(4,0){6}{\line(1,0){2}} 
\put(90,0){\makebox(0,0){$>$}}
\put(-2,19){\makebox(0,0)[t]{$0$}}
\put(-2,-11){\makebox(0,0)[t]{$1$}}
\put(20,-6){\makebox(0,0)[t]{$2$}}
\put(80,-6){\makebox(0,0)[t]{$n\!\! -\!\! 1$}}
\put(100,-7.8){\makebox(0,0)[t]{$n$}}
}
\end{picture}
(a_{ij})_{0\le i,j \le n}= 
\begin{pmatrix}
2 & 0 & -1  & \cdots & 0  & 0 
\\
0 & 2 & -1   & \cdots  & &  0
\\
-1 & -1 & 2 & \cdots & & \vdots
\\
\vdots   &  & \ddots & \ddots &  \ddots  & \vdots \\
0   &  & &  &  2 & -1 \\
0 & 0  &  \cdots &  &  -2 & 2 
\end{pmatrix}.
\end{align*}

The spin representation on $V$ is given by 
\begin{alignat}{3}
e_0|\alb \rangle 
&= z^2|\alb+{\bf e}_1+{\bf e}_2\rangle,
&\quad
f_0|\alb \rangle 
&= z^{-2}|\alb-{\bf e}_1-{\bf e}_2\rangle,
&\quad
k_0|\alb \rangle 
&= p^{2(\alpha_1+\alpha_2-1)}|\alb\rangle,
\label{B:re1}\\
e_j|\alb \rangle 
&= |\alb-{\bf e}_j+{\bf e}_{j+1}\rangle, 
&\quad
f_j|\alb \rangle 
&= |\alb+{\bf e}_j-{\bf e}_{j+1}\rangle,
&\quad
k_j |\alb \rangle
&= p^{2(\alpha_{j+1}-\alpha_{j})}|\alb\rangle\;\;\;(0<j<n),
\\
e_n|\alb \rangle 
&= |\alb-{\bf e}_n\rangle,
&\quad
f_n|\alb \rangle 
&= |\alb+{\bf e}_n\rangle,
&\quad
k_n|\alb \rangle 
&= p^{1-2\alpha_n}|\alb\rangle.
\label{B:re2}
\end{alignat}

\subsection{Onsager algebra $O_p(B^{(1)}_n)$ and the classical part $O_p(B_n)$}
\label{ss:oab}
The algebra $O_p(B^{(1)}_n)$ is generated by $\bt_0,\ldots, \bt_n$
obeying the relations (\ref{bbre1})--(\ref{bbre3}).
The classical part of $B^{(1)}_{n}$ without the vertex 0 is $B_n$.
Thus the subalgebra of $O_p(B^{(1)}_{n})$ generated by 
$\bt_1,\ldots, \bt_n$ agrees with the Onsager algebra $O_p(B_n)$
introduced in Section \ref{ss:oad2}.

\subsection{Representation $\pi^{2,1}_{2,1}$}

The representation $\pi^{2,1}_{2,1}$ of $O_p(B^{(1)}_{n})$ 
on $V$ in (\ref{Vdef}) 
is obtained by the composition 
\begin{align}\label{p2121}
\pi^{2,1}_{2,1}:
O_p(B^{(1)}_{n}) \hookrightarrow U_p(B^{(1)}_{n})  
\rightarrow \mathrm{End} V,
\end{align}
where the latter is the spin representation (\ref{B:re1})-(\ref{B:re2}) 
and the former embedding (\ref{bg0})--(\ref{drk}) reads as
\begin{align}
\bt_i &= f_i +p^{2}k^{-1}_ie_i 
- \frac{1}{t^2+t^{-2}}k^{-1}_i
\quad (0 \le  i <n),
\label{re14}\\
\bt_n 
&= f_n + p^{} k^{-1}_ne_n 
-\I \epsilon\mu\frac{t-t^{-1}}{t^2+t^{-2}}k^{-1}_n.
\label{re15}
\end{align}
This corresponds to  \cite[eqs.(167),(170)-(171)]{KOY2} with 
$s=s' = q^\hf$ according to \cite[eq.(96)]{KOY2}.
These generators are represented as local Hamiltonians:
\begin{align}
b_0 &= z^2\sigma^+_1\sigma^+_2 + z^{-2}\sigma^-_1\sigma^-_2
-\frac{t^2+t^{-2}}{4}\sigma^z_1\sigma^z_2 
+ \frac{t^2-t^{-2}}{4}(\sigma^z_1+\sigma^z_2)
+\frac{(t^2-t^{-2})^2}{4(t^2+t^{-2})},
\\
b_i & = \sigma^+_i\sigma^-_{i+1}+\sigma^-_i\sigma^+_{i+1}
-\frac{t^2+t^{-2}}{4}\sigma^z_i\sigma^z_{i+1}-
\frac{t^2-t^{-2}}{4}(\sigma^z_i-\sigma^z_{i+1})
+\frac{(t^2-t^{-2})^2}{4(t^2+t^{-2})}\quad (0<i<n),
\\
b_n &= \sigma^x_n + \mu\frac{t-t^{-1}}{2}\sigma^z_n-
\mu\frac{(t-t^{-1})(t^2-t^{-2})}{2(t^2+t^{-2})}.
\end{align}
They commute with the $K$ matrix (\ref{ktd}) up to $z$ as
\cite{KOY2}:
\begin{align}\label{kb2121}
\tilde{K}_{2,1}(z) b_i  = (b_i|_{z\rightarrow z^{-1}})\tilde{K}_{2,1}(z)
\qquad (0 \le i \le n).
\end{align} 

Set 
\begin{equation}\label{h21}
\begin{split}
H_{2,1}(z) &= b_0 + b_1 + 2(b_2+\cdots + b_{n-1}) -\mu(t+t^{-1}) b_n,
\\
&= z^2\sigma^+_1\sigma^+_2 + z^{-2}\sigma^-_1\sigma^-_2+
2\sum_{i=1}^{n-1}
\Bigl((1-{\textstyle \frac{1}{2}}\delta_{i,1})(
\sigma^+_i\sigma^-_{i+1}+\sigma^-_i\sigma^+_{i+1})
-\frac{t^2+t^{-2}}{4}\sigma^z_i\sigma^z_{i+1}
\Bigr)
\\
&-\mu(t+t^{-1})\sigma^x_n + 2n \Gamma.
\end{split}
\end{equation}
It satisfies $\sigma^x H_{2,1}(z) \sigma^x = H_{2,1}(z^{-1})$.
Therefore (\ref{kb2121}) and (\ref{kve}) lead to the commutativity
\begin{align}
[K^\vee_{2,1}(z), H_{2,1}(z)] = 0.
\end{align}

\subsection{Representation $\pi^{2,1}_{2,2}$}
The representation $\pi^{2,1}_{2,2}$ of $O_p(B^{(1)}_{n})$ 
on $V$ in (\ref{Vdef}) 
is obtained by the composition 
\begin{align}\label{p2122}
\pi^{2,1}_{2,2}:
O_p(B^{(1)}_{n}) \hookrightarrow U_p(B^{(1)}_{n})  
\rightarrow \mathrm{End} V,
\end{align}
where the latter is the spin representation (\ref{B:re1})-(\ref{B:re2}) 
and the former embedding (\ref{bg0})--(\ref{drk}) reads as
\begin{align}
\bt_i &= f_i +p^{2}k^{-1}_ie_i 
+ \frac{1}{q+q^{-1}}k^{-1}_i
\quad (0 \le  i <n),
\label{re16}\\
\bt_n 
&= f_n + p^{} k^{-1}_ne_n.
\label{re17}
\end{align}
This corresponds to \cite[eqs.(167), (180)-(181)]{KOY2}
$s=s' = q^\hf$ according to \cite[eq.(96)]{KOY2}.
These generators are represented as local Hamiltonians:
\begin{align}
b_0 &= z^2\sigma^+_1\sigma^+_2 + z^{-2}\sigma^-_1\sigma^-_2
+\frac{q+q^{-1}}{4}\sigma^z_1\sigma^z_2 
- \frac{q-q^{-1}}{4}(\sigma^z_1+\sigma^z_2)
-\frac{(q-q^{-1})^2}{4(q+q^{-1})},
\\
b_i & = \sigma^+_i\sigma^-_{i+1}+\sigma^-_i\sigma^+_{i+1}
+\frac{q+q^{-1}}{4}\sigma^z_i\sigma^z_{i+1}
+\frac{q-q^{-1}}{4}(\sigma^z_i-\sigma^z_{i+1})
-\frac{(q-q^{-1})^2}{4(q+q^{-1})}\;\;\;(0<i<n),
\\
b_n &= \sigma^x_n.
\end{align}
They commute with the $K$ matrix (\ref{ktd}) up to $z$ as
\cite{KOY2}:
\begin{align}\label{kb2122}
\tilde{K}_{2,2}(z) b_i = (b_i|_{z\rightarrow z^{-1}})\tilde{K}_{2,2}(z)
\qquad (0 \le i \le n).
\end{align}

\section{$O_p(\tilde{B}^{(1)}_n)$ Hamiltonians}\label{sec:bt}

\subsection{$U_p(\tilde{B}^{(1)}_n)$ and spin representation}
\label{ss:opbt}
The Dynkin diagram and the Cartan matrix are given by
\begin{align*}
\begin{picture}(200,50)
\put(40,0){
\put(28,24){$\mathfrak{g}^{1,2}=\tilde{B}^{(1)}_{n}$}
\put(93,14){\circle{6}}\put(93,-14){\circle{6}}
\multiput(0,0)(20,0){2}{\circle{6}}
\put(80,0){\circle{6}}
\put(23,0){\line(1,0){14}}
\put(63,0){\line(1,0){14}}
\put(82,3){\line(1,1){9}}\put(82,-3){\line(1,-1){9}}
\multiput(2.85,-1)(0,2){2}{\line(1,0){14.3}} 
\multiput(39,0)(4,0){6}{\line(1,0){2}} 
\put(10,0){\makebox(0,0){$<$}}
\put(108,18){\makebox(0,0)[t]{$n\!\! -\!\! 1$}}
\put(0,-6){\makebox(0,0)[t]{$0$}}
\put(20,-6){\makebox(0,0)[t]{$1$}}
\put(71,-6){\makebox(0,0)[t]{$n\!\! -\!\! 2$}}
\put(104,-12){\makebox(0,0)[t]{$n$}}
}
\end{picture}
(a_{ij})_{0\le i,j \le n}= 
\begin{pmatrix}
2 & -2 &    \cdots & \cdots  & 0 & 0
\\
-1 & 2 &    \cdots  & &  & 0
\\
\vdots  & \ddots &  \ddots & \ddots & & \vdots \\
\vdots     & &  &  2 & -1 & -1\\
0 &  &    &  -1 & 2 & 0
\\
0  & 0  & \cdots & -1 & 0 & 2
\end{pmatrix}.
\end{align*}

The spin representation on $V$ is given by 
\begin{alignat}{3}
e_0|\alb \rangle 
&= z|\alb+{\bf e}_1\rangle,
&\quad
f_0|\alb \rangle 
&= z^{-1}|\alb-{\bf e}_1\rangle,
&\quad
k_0|\alb \rangle 
&= p^{2\alpha_1-1}|\alb\rangle,
\label{bt:re1}\\
e_j|\alb \rangle 
&= |\alb-{\bf e}_j+{\bf e}_{j+1}\rangle, 
&\quad
f_j|\alb \rangle 
&= |\alb+{\bf e}_j-{\bf e}_{j+1}\rangle,
&\quad
k_j |\alb \rangle
&= p^{2(\alpha_{j+1}-\alpha_{j})}|\alb\rangle\;\;\;(0<j<n),
\\
e_n|\alb \rangle 
&= |\alb-{\bf e}_{n-1}-{\bf e}_n\rangle,
&\quad
f_n|\alb \rangle 
&= |\alb+{\bf e}_{n-1}+{\bf e}_n\rangle,
&\quad
k_n|\alb \rangle 
&= p^{2(1-\alpha_{n-1}-\alpha_n)}|\alb\rangle.
\label{bt:re2}
\end{alignat}

\subsection{Onsager algebra $O_p(\tilde{B}^{(1)}_n)$
and the classical part $O_p(D_n)$}\label{ss:oabt}
The algebra $O_p(\tilde{B}^{(1)}_n)$ is generated by $\bt_0,\ldots, \bt_n$
obeying the relations (\ref{bbre1})--(\ref{bbre3}).
The classical part of $\tilde{B}^{(1)}_{n}$ without the vertex 0 is $D_n$.
Thus the subalgebra of $O_p(\tilde{B}^{(1)}_{n})$ generated by 
$\bt_1,\ldots, \bt_n$ is the Onsager algebra for $D_n$.
We denote it by $O_p(D_n)$.

\subsection{Representation $\pi^{1,2}_{1,2}$}
The representation $\pi^{1,2}_{1,2}$ of $O_p(\tilde{B}^{(1)}_{n})$ 
on $V$ is obtained by the composition 
\begin{align}\label{p1212}
\pi^{1,2}_{1,2}:
O_p(\tilde{B}^{(1)}_{n}) \hookrightarrow U_p(\tilde{B}^{(1)}_{n})  
\rightarrow \mathrm{End} V,
\end{align}
where the latter is the spin representation (\ref{bt:re1})--(\ref{bt:re2})
and the former embedding (\ref{bg0})--(\ref{drk}) reads as
\begin{align}
\bt_0 
&= f_0 + p k_0^{-1}e_0
-\I \epsilon\mu\frac{t-t^{-1}}{t^2+t^{-2}}k^{-1}_0,
\label{re18}\\
\bt_i &= f_i +p^{2}k^{-1}_ie_i 
- \frac{1}{t^2+t^{-2}}k^{-1}_i
\quad (0 < i  \le n).
\label{re19}
\end{align}
This corresponds to \cite[eqs.(167), (174)-(175)]{KOY2}
$s=s' = q^\hf$ according to \cite[eq.(96)]{KOY2}.
These generators are represented as local Hamiltonians:
\begin{align}
b_0&= z \sigma^+_1+z^{-1}\sigma^-_1-\mu\frac{t-t^{-1}}{2}\sigma^z_1
- \mu\frac{(t-t^{-1})(t^2-t^{-2})}{2(t^2+t^{-2})},
\\
b_i & = \sigma^+_i\sigma^-_{i+1}+\sigma^-_i\sigma^+_{i+1}
-\frac{t^2+t^{-2}}{4}\sigma^z_i\sigma^z_{i+1}-
\frac{t^2-t^{-2}}{4}(\sigma^z_i-\sigma^z_{i+1})
+\frac{(t^2-t^{-2})^2}{4(t^2+t^{-2})}\quad (0<i<n),
\\
b_n &= \sigma^+_{n-1}\sigma^+_n + \sigma^-_{n-1}\sigma^-_n
-\frac{t^2+t^{-2}}{4}\sigma^z_{n-1}\sigma^z_n 
- \frac{t^2-t^{-2}}{4}(\sigma^z_{n-1}+\sigma^z_n)
+\frac{(t^2-t^{-2})^2}{4(t^2+t^{-2})}.
\end{align}
They commute with the $K$ matrix (\ref{ktd}) up to $z$ as
\cite{KOY2}:
\begin{align}\label{kb1212}
\tilde{K}_{1,2}(z) b_i  = (b_i|_{z\rightarrow z^{-1}})\tilde{K}_{1,2}(z)
\qquad (0 \le i \le n).
\end{align} 

Set
\begin{equation}\label{h12}
\begin{split}
H_{1,2}(z) &= -\mu(t+t^{-1}) b_0 + 2(b_1+ \cdots + b_{n-2}) + b_{n-1} + b_n
\\
&=-\mu(t+t^{-1})(z \sigma^+_1+z^{-1}\sigma^-_1)
+2\sum_{i=1}^{n-1} \Bigl(
(1-{\textstyle{\frac{1}{2}}}\delta_{i,n-1})(
\sigma^+_i\sigma^-_{i+1}+\sigma^-_i\sigma^+_{i+1})
-\frac{t^2+t^{-2}}{4}\sigma^z_i\sigma^z_{i+1}
\Bigr)
\\
&+\sigma^+_{n-1}\sigma^+_{n}+\sigma^-_{n-1}\sigma^-_{n}
+ 2n\Gamma.
\end{split}
\end{equation}
It satisfies $\sigma^x H_{1,2}(z) \sigma^x = H_{1,2}(z^{-1})$.
Therefore (\ref{kb1212}) and (\ref{kve}) lead to the commutativity
\begin{align}
[K^\vee_{1,2}(z), H_{1,2}(z)] = 0.
\end{align}

\subsection{Representation $\pi^{1,2}_{2,2}$}
The representation $\pi^{1,2}_{2,2}$ of $O_p(\tilde{B}^{(1)}_{n})$ 
on $V$ is obtained by the composition 
\begin{align}\label{p1222}
\pi^{1,2}_{2,2}:
O_p(\tilde{B}^{(1)}_{n}) \hookrightarrow U_p(\tilde{B}^{(1)}_{n})  
\rightarrow \mathrm{End} V,
\end{align}
where the latter is the spin representation (\ref{bt:re1})--(\ref{bt:re2}) an
and the former embedding (\ref{bg0})--(\ref{drk}) reads as
\begin{align}
\bt_0 
&= f_0 + p k_0^{-1}e_0,
\label{re20}\\
\bt_i &= f_i +p^{2}k^{-1}_ie_i 
+ \frac{1}{q+q^{-1}}k^{-1}_i
\quad (0 < i  \le n).
\label{re21}
\end{align}
This corresponds to \cite[eqs.(167), (182)-(183)]{KOY2}
$s=s' = q^\hf$ according to \cite[eq.(96)]{KOY2}.
These generators are represented as local Hamiltonians:
\begin{align}
b_0&= z \sigma^+_1+z^{-1}\sigma^-_1,
\\
b_i & = \sigma^+_i\sigma^-_{i+1}+\sigma^-_i\sigma^+_{i+1}
+\frac{q+q^{-1}}{4}\sigma^z_i\sigma^z_{i+1}
+\frac{q-q^{-1}}{4}(\sigma^z_i-\sigma^z_{i+1})
-\frac{(q-q^{-1})^2}{4(q+q^{-1})}\;\;\;(0<i<n),
\\
b_n &= \sigma^+_{n-1}\sigma^+_n + \sigma^-_{n-1}\sigma^-_n
+\frac{q+q^{-1}}{4}\sigma^z_{n-1}\sigma^z_n 
+ \frac{q-q^{-1}}{4}(\sigma^z_{n-1}+\sigma^z_n)
-\frac{(q-q^{-1})^2}{4(q+q^{-1})}.
\end{align}
They commute with the $K$ matrix (\ref{ktd}) up to $z$ as
\cite{KOY2}:
\begin{align}\label{kb1222}
\tilde{K}_{2,2}(z) b_i  = (b_i|_{z\rightarrow z^{-1}})\tilde{K}_{2,2}(z)
\qquad (0 \le i \le n).
\end{align} 

\section{$O_p(D^{(1)}_n)$ Hamiltonians}
\label{sec:D1}

\subsection{$U_p(D^{(1)}_n)$ and spin representations}
\label{ss:upd1n}
The Dynkin diagram and the Cartan matrix are given by
\begin{align*}
\begin{picture}(200,50)
\put(50,0){
\put(32,24){$\mathfrak{g}^{2,2}=D^{(1)}_n$}
\put(6,14){\circle{6}}\put(6,-14){\circle{6}}
\put(20,0){\circle{6}}
\put(80,0){\circle{6}}
\put(93,14){\circle{6}}\put(93,-14){\circle{6}}
\put(18,3){\line(-1,1){9}} \put(18,-3){\line(-1,-1){9}}
\put(23,0){\line(1,0){14}}
\multiput(39,0)(4,0){6}{\line(1,0){2}} 
\put(62.5,0){\line(1,0){14}}
\put(82,3){\line(1,1){9}}\put(82,-3){\line(1,-1){9}}
\put(-2,19){\makebox(0,0)[t]{$0$}}
\put(-2,-11){\makebox(0,0)[t]{$1$}}
\put(20,-6){\makebox(0,0)[t]{$2$}}
\put(71,-6){\makebox(0,0)[t]{$n\!\! -\!\! 2$}}
\put(108,18){\makebox(0,0)[t]{$n\!\! -\!\! 1$}}
\put(104,-12){\makebox(0,0)[t]{$n$}}
}
\end{picture}
(a_{ij})_{0\le i,j \le n}= 
\begin{pmatrix}
2 & 0 & -1  & \cdots & \cdots  & 0 & 0
\\
0 & 2 & -1   & \cdots  & &  & 0 
\\
-1 & -1 & 2 & \cdots & & & \vdots
\\
\vdots   &  & \ddots & \ddots &  \ddots & & \vdots \\
\vdots   &  & &  &  2 & -1 & -1\\
0 &  &  &  &  -1 & 2 & 0
\\
0 & 0 & \cdots  & \cdots & -1 & 0 & 2
\end{pmatrix}.
\end{align*}

There are two spin representations $V_+, V_- \subset V$ 
in (\ref{vl})  
with dimension $\dim V_\pm = 2^{n-1}$.
They are given by 
\begin{alignat}{3}
e_0|\alb \rangle 
&= z^2|\alb+{\bf e}_1+{\bf e}_2\rangle,
&\quad
f_0|\alb \rangle 
&= z^{-2}|\alb-{\bf e}_1-{\bf e}_2\rangle,
&\quad
k_0|\alb \rangle 
&= p^{2(\alpha_1+\alpha_2-1)}|\alb\rangle,
\label{d1:re1}\\
e_j|\alb \rangle 
&= |\alb-{\bf e}_j+{\bf e}_{j+1}\rangle, 
&\quad
f_j|\alb \rangle 
&= |\alb+{\bf e}_j-{\bf e}_{j+1}\rangle,
&\quad
k_j |\alb \rangle
&= p^{2(\alpha_{j+1}-\alpha_{j})}|\alb\rangle\;\;\;(0<j<n),
\\
e_n|\alb \rangle 
&= |\alb-{\bf e}_{n-1}-{\bf e}_n\rangle,
&\quad
f_n|\alb \rangle 
&= |\alb+{\bf e}_{n-1}+{\bf e}_n\rangle,
&\quad
k_n|\alb \rangle 
&= p^{2(1-\alpha_{n-1}-\alpha_n)}|\alb\rangle.
\label{d1:re2}
\end{alignat}

\subsection{Onsager algebra $O_p(D^{(1)}_{n})$ and the classical part $O_p(D_n)$}
\label{ss:oad1}
The algebra $O_p(D^{(1)}_{n})$ is generated by $\bt_0,\ldots, \bt_n$
obeying the relations (\ref{bbre1})--(\ref{bbre3}).
The classical part of $D^{(1)}_{n}$ without the vertex 0 is $D_n$.
Thus the subalgebra of $O_p(D^{(1)}_{n})$ generated by 
$\bt_1,\ldots, \bt_n$ agrees with the Onsager algebra $O_p(D_n)$
introduced in Section \ref{ss:oabt}.

\subsection{Representation $\pi^{2,2}_{2,2}$}
The representation $\pi^{2,2}_{2,2}$ of $O_p(D^{(1)}_{n})$ 
on $V_\pm$ is obtained by the composition 
\begin{align}\label{p2222}
\pi^{2,2}_{2,2}:
O_p(D^{(1)}_{n}) \hookrightarrow U_p(D^{(1)}_{n})  
\rightarrow \mathrm{End} V_+ \oplus \mathrm{End} V_-,
\end{align}
where the latter is the spin representation (\ref{d1:re1})-(\ref{d1:re2}) 
and the former embedding (\ref{bg0})--(\ref{drk}) reads as
\begin{align}
\bt_i &= f_i +p^2k^{-1}_ie_i 
+ \frac{1}{q+q^{-1}}k^{-1}_i
\quad (0 \le i \le n).
\label{re13}
\end{align}
This corresponds to  \cite[eqs.(167), (178)-(179)]{KOY2} with 
$s=s' = q^\hf$ according to \cite[eq.(96)]{KOY2}.
These generators are represented as local Hamiltonians:
\begin{align}
b_0 &= z^2\sigma^+_1\sigma^+_2 + z^{-2}\sigma^-_1\sigma^-_2
+\frac{q+q^{-1}}{4}\sigma^z_1\sigma^z_2 
- \frac{q-q^{-1}}{4}(\sigma^z_1+\sigma^z_2)
-\frac{(q-q^{-1})^2}{4(q+q^{-1})},
\label{hota1}\\
b_i & = \sigma^+_i\sigma^-_{i+1}+\sigma^-_i\sigma^+_{i+1}
+\frac{q+q^{-1}}{4}\sigma^z_i\sigma^z_{i+1}
+\frac{q-q^{-1}}{4}(\sigma^z_i-\sigma^z_{i+1})
-\frac{(q-q^{-1})^2}{4(q+q^{-1})}\;\;\;(0<i<n),
\\
b_n &= \sigma^+_{n-1}\sigma^+_n + \sigma^-_{n-1}\sigma^-_n
+\frac{q+q^{-1}}{4}\sigma^z_{n-1}\sigma^z_n 
+ \frac{q-q^{-1}}{4}(\sigma^z_{n-1}+\sigma^z_n)
-\frac{(q-q^{-1})^2}{4(q+q^{-1})}.
\label{hota2}
\end{align}
They commute with the $K$ matrix (\ref{ktd}) up to $z$ as
\cite{KOY2}:
\begin{align}\label{kb2222}
\tilde{K}_{2,2}(z) b_i  = (b_i|_{z\rightarrow z^{-1}})\tilde{K}_{2,2}(z)
\qquad (0 \le i \le n).
\end{align} 

Set
\begin{equation}\label{h22}
\begin{split}
H_{2,2}(z) &= b_0+b_1+2(b_2+\cdots + b_{n-2}) + b_{n-1} + b_n
\\
&= z^2\sigma^+_1\sigma^+_2 + z^{-2}\sigma^-_1\sigma^-_2
+
2\sum_{i=1}^{n-1}\Bigl(
(1-{\textstyle \frac{1}{2}}\delta_{i,1}-{\textstyle \frac{1}{2}}\delta_{i,n-1})
(\sigma^+_i\sigma^-_{i+1}+\sigma^-_i\sigma^+_{i+1})
+ \frac{q+q^{-1}}{4}\sigma^z_i\sigma^z_{i+1}
\Bigr) 
\\
&+ \sigma^+_{n-1}\sigma^+_n + \sigma^-_{n-1}\sigma^-_n +(2n-2)\Gamma.
\end{split}
\end{equation}
It satisfies $\sigma^x H_{2,2}(z) \sigma^x = H_{2,2}(z^{-1})$.
Therefore (\ref{kb2222}) and (\ref{kve}) lead to the commutativity
\begin{align}
[K^\vee_{2,2}(z), H_{2,2}(z)] = 0.
\end{align}

\section{Spectral decomposition of $K_{1,1}$ and $K_{2,1}$ by $O_p(B_n)$}
\label{sec:spdb}

The Onsager algebra $O_p(B_n)$ defined in Section \ref{ss:oad2}
shows up either as the classical part of $O_p(D^{(2)}_{n+1})$
or $O_p(B^{(1)}_n)$.
Consider the resulting representations of $O_p(B_n)$ constructed as
\begin{align}
O_p(B_n)  &\hookrightarrow  O_p(D^{(2)}_{n+1}) 
\overset{\pi^{1,1}_{1,1}}{\longrightarrow}
\mathrm{End} V,
\\
O_p(B_n)  &\hookrightarrow  O_p(D^{(2)}_{n+1}) 
\overset{\pi^{1,1}_{2,1}}{\longrightarrow}
\mathrm{End} V,
\\
O_p(B_n)  &\hookrightarrow  O_p(B^{(1)}_{n}) 
\overset{\pi^{2,1}_{2,1}}{\longrightarrow}
\mathrm{End} V.
\end{align}
From the definitions (\ref{p1111}), (\ref{p1121}) and (\ref{p2121}),
they actually yield the same representation
\begin{align}
b_i & = \sigma^+_i\sigma^-_{i+1}+\sigma^-_i\sigma^+_{i+1}
-\frac{t^2+t^{-2}}{4}\sigma^z_i\sigma^z_{i+1}-
\frac{t^2-t^{-2}}{4}(\sigma^z_i-\sigma^z_{i+1})
+\frac{(t^2-t^{-2})^2}{4(t^2+t^{-2})}\;\;\;(1\le i<n),
\\
b_n &= \sigma^x_n + \mu\frac{t-t^{-1}}{2}\sigma^z_n-
\mu\frac{(t-t^{-1})(t^2-t^{-2})}{2(t^2+t^{-2})}.
\end{align}
We denote this by $\pi^1_1: O_p(B_n) \rightarrow \mathrm{End}V$. 

The relation (\ref{kb1111}) with $i \neq 0$ tells that 
$\pi^1_1(O_p(B_{n}))$ commutes with $\tilde{K}_{1,1}(z)$.
Similarly, the both (\ref{kb1121}) and (\ref{kb2121}) imply 
that it also commutes with $\tilde{K}_{2,1}(z)$.
We summarize these facts as 
\begin{align}
[\tilde{K}_{1,1}(z), \pi^1_1(O_p(B_{n}))]=0,\quad
[\tilde{K}_{2,1}(z), \pi^1_1(O_p(B_{n}))]=0.
\end{align}
In other words, 
there are at least two affinizations that are compatible with 
the classical Onsager algebra symmetry $O_p(B_n)$ in the 
representation $\pi^1_1$ under consideration. 

The representations
$\pi^{1,1}_{1,1}, \pi^{1,1}_{2,1}$  of $O_p(D^{(2)}_{n+1})$ 
and $\pi^{2,1}_{2,1}$ of $O_p(B^{(1)}_n)$ on $V$ are irreducible  \cite{KOY2}.
On the other hand $V$ is no longer irreducible with respect to their common 
classical subalgebra $O_p(B_n)$.
The $K$ matrices should be scalar 
on each irreducible component.
We conjecture that there are irreducible 
$O_p(B_n)$ modules $X_0, X_1, \ldots, X_n$ 
allowing the joint spectral decomposition of 
$K_{1,1}(z)$ and $K_{2,1}(z)$ as follows:
\begin{align}
V &= X_0 \oplus X_1 \oplus \cdots \oplus X_n,
\quad \dim X_l = \binom{n}{l},
\\
K_{1,1}(z) &= 
\bigoplus_{0 \le l \le n}
z^{n-2l}\frac{(-q^{n+1-2l}z;q)_\infty(-qz^{-1};q)_\infty}
{(-qz;q)_\infty(-q^{n+1-2l}z^{-1};q)_\infty}\mathrm{id}_{X_l},
\label{ks11}\\
K_{2,1}(z) &= \bigoplus_{0 \le l \le n}
z^{n-2l}\frac{(-q^{2n+3-4l}z^2;q^4)_\infty(-q^3z^{-2};q^4)_\infty}
{(-q^3z^2;q^4)_\infty(-q^{2n+3-4l}z^{-2};q)_\infty}\mathrm{id}_{X_l}
\quad(n \;\text{even}),
\\
&= \bigoplus_{0 \le l \le n}
z^{n+1-2l}\frac{(-q^{2n+3-4l}z^2;q^4)_\infty(-qz^{-2};q^4)_\infty}
{(-qz^2;q^4)_\infty(-q^{2n+3-4l}z^{-2};q^4)_\infty}\mathrm{id}_{X_l}
\quad(n \;\text{odd}).
\end{align} 
Here $\mathrm{id}_{X_l}$ denotes the 
orthonormal projector 
$\mathrm{id}_{X_l}\mathrm{id}_{X_{l'}}
= \delta_{l,l'}\mathrm{id}_{X_l}$.
Similar notations will also be used in the sequel.
(Note that $\tilde{K}_{k,k'}(z)$ 
and $K_{k,k'}(z)$ possess the same spectrum due to (\ref{ktd}).)

As an example (\ref{ks11}) means
\begin{align}\label{sept}
K_{1,1}(z) 
&= 
\bigoplus_{l \ge [\frac{n+1}{2}]}
\prod_{j=1}^{2l-n}\frac{q^j+z}{1+q^jz}\mathrm{id}_{X_l}
\oplus 
\bigoplus_{l \le [\frac{n-1}{2}]}
\prod_{j=1}^{n-1-2l}\frac{q^j+z}{1+q^jz}\mathrm{id}_{X_l},
\end{align} 
where $[x]$ stands for the largest integer not exceeding $x$.

\section{Spectral decomposition of $K_{1,2}$ and $K_{2,2}$ by $O_p(D_n)$}
\label{sec:spdd}

The Onsager algebra $O_p(D_n)$ defined in Section \ref{ss:oabt}
shows up either as the classical part of 
$O_p(\tilde{B}^{(1)}_n)$ or $O_p(D^{(1)}_n)$.
Consider the resulting representations of $O_p(D_n)$ constructed as 
\begin{align}
O_p(D_n)  &\hookrightarrow  O_p(\tilde{B}^{(1)}_n)
\overset{\pi^{1,2}_{1,2}}{\longrightarrow}
\mathrm{End} V,
\\
O_p(D_n)  &\hookrightarrow  O_p(\tilde{B}^{(1)}_n)
\overset{\pi^{1,2}_{2,2}}{\longrightarrow}
\mathrm{End} V,
\\
O_p(D_n)  &\hookrightarrow  O_p(D^{(1)}_{n}) 
\overset{\pi^{2,2}_{2,2}}{\longrightarrow}
\mathrm{End} V_+ \oplus \mathrm{End} V_-.
\end{align}
For the definition of $V_\pm$, see (\ref{vl}).
From (\ref{p1212}), (\ref{p1222}) and (\ref{p2222}), 
they actually yield the same representation 
\begin{align}
b_i & = \sigma^+_i\sigma^-_{i+1}+\sigma^-_i\sigma^+_{i+1}
+\frac{q+q^{-1}}{4}\sigma^z_i\sigma^z_{i+1}
+\frac{q-q^{-1}}{4}(\sigma^z_i-\sigma^z_{i+1})
-\frac{(q-q^{-1})^2}{4(q+q^{-1})}\;\;\;(1\le i<n),
\\
b_n &= \sigma^+_{n-1}\sigma^+_n + \sigma^-_{n-1}\sigma^-_n
+\frac{q+q^{-1}}{4}\sigma^z_{n-1}\sigma^z_n 
+ \frac{q-q^{-1}}{4}(\sigma^z_{n-1}+\sigma^z_n)
-\frac{(q-q^{-1})^2}{4(q+q^{-1})}.
\end{align}
Obviously this defines the representation either 
on $V_+$ or $V_-$ separately.
We denote them by 
$\pi^2_{2,\pm}: O_p(D_n) \rightarrow \mathrm{End} V_\pm$
and their direct sum by 
$\pi^2_2: O_p(D_n) \rightarrow \mathrm{End} V$.
 
The $K$ matrix $\tilde{K}_{1,2}(z)$ does not preserve $V_+$ and $V_-$ individually.
However the relation(\ref{kb1212}) tells that it 
commutes with $\pi^{2}_2(O_p(D_n))$. 
On the other hand, 
$\tilde{K}_{2,2}(z)$ 
maps $V_\pm$ to $V_{\pm (-1)^n}$ as seen in (\ref{askB}).
Then  
(\ref{kb1222}) and (\ref{kb2222}) imply 
\begin{align}
[\tilde{K}_{1,2}(z), \pi^{2}_2(O_p(D_n))]=0,
\quad 
\pi^{2}_{2, \pm (-1)^n}\!(O_p(D_n))\tilde{K}_{2,2}(z) = 
\tilde{K}_{2,2}(z)\pi^{2}_{2,\pm}(O_p(D_n)).
\end{align}

The representations $\pi^{1,2}_{1,2}, \pi^{1,2}_{2,2}$ of 
$O_p(\tilde{B}^{(1)}_n)$ on $V$ and 
$\pi^{2,2}_{2,2}$ of $O_p(D^{(1)}_n)$ on $V_\pm$ are
irreducible \cite{KOY2}.
On the other hand they are no longer irreducible with respect to their 
common classical subalgebra $O_p(D_n)$.
The $K$ matrices should be a scalar on each irreducible component.

For $n$ even, we conjecture that there are irreducible $O_p(D_n)$ modules
$Z^\pm_0, \ldots, Z^\pm_{\frac{n}{2}}$
having the properties (i) and  (ii) described below:

(i) $V_\pm$ are decomposed as
\begin{align}
V_\pm &= Z^\pm_0 \oplus \cdots \oplus Z^\pm_{\frac{n}{2}},
\quad 
\dim Z^{\pm}_l = 
\begin{cases} 
\binom{n}{l} & (0 \le l < \frac{n}{2}),\\
\frac{1}{2}\binom{n}{\frac{n}{2}} & (l = \frac{n}{2}),
\end{cases}
\end{align} 
which is consistent with $\dim V_\pm = 2^{n-1}$.

(ii) There exists a basis
$\{ \zeta^\pm_{l,i}\mid 1 \le i \le \dim Z^\pm_l\}$
of $Z^\pm_l$ in terms of which the spectral decomposition of the 
$K$ matrices is described as
\begin{align}
K_{1,2}(z)& = \bigoplus_{0 \le l \le n}
\frac{(-tz^{-1};t^4)_\infty(-t^{2n+1-4l}z;t^4)_\infty
(tz^{-1};t^4)_\infty(t^{-2n+1+4l}z;t^4)_\infty}
{(-t^{2n+1-4l}z^{-1};t^4)_\infty(-tz;t^4)_\infty
(t^{-2n+1+4l}z^{-1};t^4)_\infty(tz;t^4)_\infty}\mathrm{id}_{Y_l},
\\
K_{2,2}(z) &= 
\bigoplus_{0 \le l \le \frac{n}{2}}
\frac{(-t^2z^{-1};t^4)_\infty(-t^{2n+2-4l}z;t^4)_\infty
(t^2z^{-1};t^4)_\infty(t^{-2n+2+4l}z;t^4)_\infty}
{(-t^{2n+2-4l}z^{-1};t^4)_\infty(-t^2z;t^4)_\infty
(t^{-2n+2+4l}z^{-1};t^4)_\infty(t^2z;t^4)_\infty}
(\mathrm{id}_{Z^+_l} \oplus \mathrm{id}_{Z^+_l}),
\label{kwg1}
\end{align}
where the spaces $Y_0,\ldots, Y_n$ are  given by 
\begin{align}
Y_l & = \bigoplus_{1 \le i \le \binom{n}{l}}
\C (\zeta^+_{l,i}+ \zeta^-_{l,i})\quad
(0 \le l < \frac{n}{2}),
\label{yd1}\\
Y_l&= \bigoplus_{1 \le i \le \frac{1}{2}\binom{n}{\frac{n}{2}}}
(\C \zeta^+_{\frac{n}{2},i} + \C \zeta^-_{\frac{n}{2},i})
\quad( l= \frac{n}{2}),
\label{yd2}\\
Y_l&= \bigoplus_{1 \le i \le \binom{n}{l}}
\C (\zeta^+_{n-l,i}- \zeta^-_{n-l,i})\quad
(\frac{n}{2} < l \le n).
\label{yd3}
\end{align}
Thus the following relations hold:
\begin{align}
Y_{\frac{n}{2}} \cap V_\pm = Z^\pm_{\frac{n}{2}},
\quad (Y_l + Y_{n-l}) \cap V_\pm = Z^\pm_l \;\; (0 \le l < \frac{n}{2}).
\label{ayami}
\end{align}

For $n$ odd, we conjecture that there are irreducible $O_p(D_n)$ modules
$Z^\pm_0, \ldots, Z^\pm_{\frac{n-1}{2}}$
having the properties (iii) and  (iv) described below:

(iii) $V_\pm$ are decomposed as
\begin{align}
V_\pm &= Z^\pm_0 \oplus \cdots \oplus Z^\pm_{\frac{n-1}{2}},
\quad 
\dim Z^{\pm}_l = 
\binom{n}{l},
\end{align} 
which is consistent with $\dim V_\pm = 2^{n-1}$.

(iv) There exists a basis
$\{ \zeta^\pm_{l,i}\mid 1 \le i \le \dim Z^\pm_l\}$
of $Z^\pm_l$ in term of which the spectral decomposition 
of the $K$ matrices is described as
\begin{align}
K_{1,2}(z) &= \bigoplus_{0 \le l \le n}
z\frac{(-tz^{-1};t^4)_\infty(-t^{2n+3-4l}z;t^4)_\infty
(tz^{-1};t^4)_\infty(t^{-2n+3+4l}z;t^4)_\infty}
{(-t^{2n+3-4l}z^{-1};t^4)_\infty(-tz;t^4)_\infty
(t^{-2n+3+4l}z^{-1};t^4)_\infty(tz;t^4)_\infty}\mathrm{id}_{Y_l},
\\
K_{2,2}(z) &= \bigoplus_{0 \le l \le \frac{n-1}{2}}
\frac{-t(-t^4z^{-1};t^4)_\infty(-t^{2n+2-4l}z;t^4)_\infty
(t^4z^{-1};t^4)_\infty(t^{-2n+2+4l}z;t^4)_\infty}
{z(-t^{2n+2-4l}z^{-1};t^4)_\infty(-z;t^4)_\infty
(t^{-2n+2+4l}z^{-1};t^4)_\infty(z;t^4)_\infty}
(P^+_l \oplus P^-_l).
\label{kwg2}
\end{align}
Here the spaces $Y_0, \ldots, Y_n$ are 
given again by (\ref{yd1}) and (\ref{yd3}),
hence the latter relation in (\ref{ayami}) is valid.
The operators $P^\pm_l$ are defined by
\begin{align}
P^\pm_l \zeta^\pm_{l',i} = \delta_{l,l'} \zeta^\mp_{l,i},
\quad
P^\pm_l \zeta^\mp_{l',i} = 0,
\end{align}
giving isomorphism $Z^\pm_l \rightarrow Z^\mp_l$.
We note that the eigenvalues appearing in 
(\ref{kwg1}) and (\ref{kwg2}) are actually even functions of $z$
as with $K_{2,2}(z)$.

\section{Summary}\label{sec:sum}

In this paper we have pointed out that the  
generators of the Onsager algebras
$O_p(A^{(1)}_{n-1})$ in the fundamental representations and 
$O_p(D^{(2)}_{n+1})$, 
$O_p(B^{(1)}_n)$,
$O_p(\tilde{B}^{(1)}_n)$, 
$O_p(D^{(1)}_n)$ 
in the spin representations are naturally regarded as  
local Hamiltonians of XXZ type spin chains 
involving various boundary terms reflecting the relevant Dynkin diagrams. 
The reflection $K$ matrices due to the matrix product construction \cite{KP} 
are shown to serve as symmetry operators of these Hamiltonians.
The spectra of the latters are yet to be analyzed in general. 
We have given the spectral decomposition of the $K$ matrices 
with respect to the classical part of the Onsager algebras conjecturally.
They exhibit an intriguing structure which deserves further investigations
from the viewpoint of the representation theory of Onsager algebras.  
We have also included a proof of 
Theorem \ref{th:bv}, which was formulated as a conjecture in \cite{KP}
and played a key role in the matrix product construction there. 

Let us remark a related result from \cite{UI}, 
where a family of mutually commuting Hamiltonians of the form 
\begin{align}
I_m = \kappa_0 S^{(m)}_0 + \kappa_1 S^{(m)}_1 + \cdots + 
\kappa_{n'}S^{(m)}_{n'},
\quad S^{(m)}_i \in O_p(\mathfrak{g}),\quad S^{(0)}_i = \bt_i\quad (m \ge 0)
\end{align}
were constructed for $\mathfrak{g}=A^{(1)}_{n-1}$ (hence $n'=n-1$) with $n\ge 3$ 
and $p=1$.
Here $\kappa_0,\ldots, \kappa_{n'}$ are {\em free} parameters. 
(A slightly more general one is given in \cite[eq.(2.44)]{BCP}.)
Our $H_{\mathrm{tr}}(z)= b_0 + \cdots + b_{n-1}$ 
in (\ref{kbtr}) and (\ref{han}) formally 
correspond to a $p$-analogue of the representation of $I_0$ on $V$ with 
$\forall \kappa_i=1$.
It is an interesting problem to construct a $p$-analogue of 
$I_m$ within $O_p(\mathfrak{g})$ for general $\mathfrak{g}$.

\appendix

\section{Proof of Proposition \ref{pr:kk}}\label{app:kcom}

Define 
\begin{align}
K(z)_{\alb}^{\beb} = 
\mathrm{Tr}(z^{\bf h} K_{\alpha_1}^{\beta_1} \cdots 
K_{\alpha_n}^{\beta_n} ), 
\quad
K(z)|\alb\rangle = 
\sum_{\beb \in \{0,1\}^n}K(z)_{\alb}^{\beb}|\beb\rangle,
\end{align}
which are just (\ref{mpktr}) and (\ref{ktr1}) 
without an overall scalar
for simplicity.
In order to describe the elements of $K(z)K(w)$, 
we prepare two copies of (\ref{kop})
and their product:
\begin{align}\label{ikasan}
&L_i = 
\begin{pmatrix}
\ap_i & -q \ok_i\\ \ok_i & \am_i
\end{pmatrix},
\quad
\begin{pmatrix}
M^0_0   & M^0_1 \\ M^1_0 & M^1_1
\end{pmatrix}
=
L_1 \cdot L_2 = 
\begin{pmatrix}
\ap_1 \ap_2 -q \ok_1\ok_2 & -q (\ap_1\ok_2+\ok_1 \am_2)
\\
\ok_1 \ap_2 + \am_1 \ok_2 & 
\am_1 \am_2-q \ok_1 \ok_2
\end{pmatrix},
\end{align}
where $i=1,2$ and $\cdot$ signifies the usual product as 2 by 2 matrices.
We will also use the copies $\h_1, \h_2$ of 
the number operator $\h$ defined after (\ref{qbos}).
Operators with different indices are commutative 
as they act on different $q$-boson Fock spaces.

By the definition the matrix element of $K(z)K(w)$ is expressed as
\begin{align}\label{jmp}
&(K(z)K(w) )_{\alpha_1,\ldots, \alpha_n}^{\beta_1,\ldots, \beta_n}
= \mathrm{Tr}_{12}\left(
z^{\h_1}w^{\h_2}  M^{\beta_1}_{\alpha_1}
\cdots M^{\beta_n}_{\alpha_n}\right),
\end{align}
where the trace extends over the two $q$-boson Fock spaces 1 and 2.

Let $r$ be the exchange operator of the two $q$-bosons:
\begin{align}
r^2 = 1,\quad r \,\apm_i = \apm_{3-i} r,\quad 
r\, \ok_i = \ok_{3-i} r,\quad 
r \h_i = \h_{3-i} r\qquad (i=1,2).
\label{pet0}
\end{align}
One can easily check the following relations for any $\alpha, \beta \in \{0,1\}$:
\begin{align}
&r M^{\beta}_{\alpha} = (-q)^{\alpha-\beta}M^{\alpha}_{\beta}r,
\label{pet1}\\
&M^0_1M^1_0 = M^1_0M^0_1,\quad
M^{\alpha}_{1-\alpha} M^0_0 = q M^0_0 M^{\alpha}_{1-\alpha},
\quad 
M^{\alpha}_{1-\alpha} M^1_1 = q^{-1} M^1_1 M^{\alpha}_{1-\alpha}.
\label{pet2}
\end{align}
Note that the relation (\ref{pet2}) 
is also satisfied by the elements of $L_i$ (\ref{ikasan}) for each $i=1,2$.
The product $L_1\cdot L_2$ preserves the relation
because it coincides with the coproduct of the 
$q$-oscillator representation $\pi_i$ of the quantized coordinate ring 
$A_q(\mathrm{SL}_n)$ in \cite[eqs.(2.3)--(2.6)]{KO}.

Insert $1=r^2$ anywhere in the trace of (\ref{jmp})
and let one the $r$'s encircle the whole array once 
using (\ref{pet0}) and (\ref{pet1}).
The result gives
\begin{align}
&\mathrm{Tr}_{12}\left(
z^{\h_1}w^{\h_2}  M^{\beta_1}_{\alpha_1}
\cdots M^{\beta_n}_{\alpha_n}\right)
= \mathrm{Tr}_{12}\left(
w^{\h_1}z^{\h_2}  M_{\beta_1}^{\alpha_1}
\cdots M_{\beta_n}^{\alpha_n}\right)(-q)^{|\alb|- |\beb|},
\end{align}
where the symbol $|\alb|$ is defined in (\ref{aoda}).
From (\ref{ktrd}) we know 
$(K(z)K(w) )_{\alpha_1,\ldots, \alpha_n}^{\beta_1,\ldots, \beta_n}=0$
unless $|\alb| = |\beb|$.
Thus the factor $(-q)^{|\alb|- |\beb|}$ in the above can be removed,
leading to
\begin{align}\label{pet3}
(K(z)K(w) )_{\alpha_1,\ldots, \alpha_n}^{\beta_1,\ldots, \beta_n}
=
(K(w)K(z) )^{\alpha_1,\ldots, \alpha_n}_{\beta_1,\ldots, \beta_n}.
\end{align} 

Next consider the expression (\ref{jmp}) again.
Under the assumption $|\alb| = |\beb|$, the number 
of $M^1_0$ and $M^0_1$ in the trace is equal, which we denote by $m$.
Then by means of (\ref{pet2}) one can send 
$M^1_0$ and $M^0_1$ to the left to rewrite (\ref{jmp}) uniquely in the form
\begin{align}
(K(z)K(w) )_{\alpha_1,\ldots, \alpha_n}^{\beta_1,\ldots, \beta_n}
= q^{\Phi}\,\mathrm{Tr}_{12}\left(
z^{\h_1}w^{\h_2}  (M^1_0 M^0_1)^m N_1\cdots N_{n-2m}\right),
\end{align}
where 
$N_i = M^0_0 \;\,\text{or}\;\, M^1_1$ are in the original order 
and $\Phi$ is some integer.
Starting from $(K(z)K(w) )^{\alpha_1,\ldots, \alpha_n}_{\beta_1,\ldots, \beta_n}$,
the same rewriting procedure leads to the identical expression 
due to (\ref{pet2}). 
Thus we find 
\begin{align}\label{pet4}
(K(z)K(w) )_{\alpha_1,\ldots, \alpha_n}^{\beta_1,\ldots, \beta_n}
= (K(z)K(w) )^{\alpha_1,\ldots, \alpha_n}_{\beta_1,\ldots, \beta_n}.
\end{align}
Combining (\ref{pet4}) with (\ref{pet3}) we conclude
\begin{align}
(K(w)K(z) )^{\alpha_1,\ldots, \alpha_n}_{\beta_1,\ldots, \beta_n}
= (K(z)K(w) )^{\alpha_1,\ldots, \alpha_n}_{\beta_1,\ldots, \beta_n},
\end{align}
which completes a proof of (\ref{com}).

\section{Proof of the invariance of boundary vectors under 3D $K$ matrix}\label{app:A}

The matrix product construction of the reflection $K$ matrices 
$K_{k,k'}(z)$  in \cite{KP} was based on the fact 
that certain boundary vectors remain invariant 
under the action of the {\em 3D $K$} matrix $\mathscr{K}$
which is the intertwiner of quantized coordinate ring $A_q(\mathrm{Sp}_4)$ \cite{KO}.
See \cite[eq.(78)]{KP}.
In this appendix we prove this crucial 
property which had been left as a conjecture in 
\cite{KP}, thereby completing the 3D approach there.
For simplicity we shall concentrate 
on the latter relation in \cite[eq.(78)]{KP} on the ket-vectors.
The former relation corresponding to the bra-vector version follows from it 
by an argument similar to the proof of \cite[Prop.2.4]{KO}.
We leave an detailed description of the 3D $K$ matrix
to the original work \cite{KO}.
A quick exposition is available in \cite[Sec.3.2]{KP}.

Let $F_{q^2}$ be the Fock space  
obtained by formally replacing $q$ by $q^2$ in $F_q$ in Section \ref{ss:kmat}.
The $q^2$-boson operators are denoted by $\Apm, \OK$, i,e,,
\begin{alignat}{3}
\Am|m\rangle &= (1-q^{4m})|m-1\rangle,
\quad & \Ap|m\rangle &= |m+1\rangle,
\quad & \OK|m\rangle &=q^{2m}|m\rangle.
\end{alignat}
We introduce the boundary vectors by 
\begin{align}
|\chi_k\rangle = \sum_{m\ge 0}\frac{|km\rangle}{(q^{2k^2};q^{2k^2})_m} \in F_{q^2}
\qquad (k=1,2),
\end{align}
which is equal to $|\eta_k\rangle$ in (\ref{xb}) with $q$ replaced by $q^2$.
Up to normalization, the vector 
$|\eta_1\rangle$ ($|\chi_1\rangle$) is 
characterized by any one of the following three conditions in the left (right) column:
\begin{alignat}{2}
(\ap-1+\ok) |\eta_1\rangle &=0,
& \qquad (\Ap-1+\OK) |\chi_1\rangle &=0,
\label{c11}\\
(\am -1-q\ok)|\eta_1\rangle &=0,
& \qquad (\Am -1-q^2\OK)|\chi_1\rangle &=0,
\label{c12}\\
(\ap-\am + (1+q)\ok)|\eta_1\rangle &=0,
& \qquad (\Ap-\Am + (1+q^2)\OK)|\chi_1\rangle &=0.
\label{c13}
\end{alignat}
Up to normalization, the vectors 
$|\eta_2\rangle$ and $|\chi_2\rangle$ are characterized by
\begin{align}\label{c2}
(\ap-\am)|\eta_2\rangle=0,\qquad 
(\Ap-\Am)|\chi_2\rangle=0.
\end{align}
Define the three boundary vectors by
\begin{align}\label{bv}
|\Xi_{r,k}\rangle = 
|\chi_r\rangle \otimes |\eta_k\rangle \otimes |\chi_r\rangle \otimes |\eta_k\rangle
\qquad ((r,k)=(1,1), (1,2), (2,2)).
\end{align}

Let $\mathscr{K} 
\in \mathrm{End}(F_{q^2}\otimes F_{q}\otimes F_{q^2}\otimes F_{q})$ 
be the 3D $K$ matrix in \cite[Th.3.4]{KO} which only depends on the parameter $q$.
It satisfies the intertwining relation
\begin{align}\label{ir}
\Delta(t_{ij}) \mathscr{K} = \mathscr{K} \Delta^{\!\mathrm{op}}(t_{ij})
\qquad (i,j \in \{1,2,3,4\}),
\end{align}
where $\Delta$ and $\Delta^{\!\mathrm{op}}$ are shorthand for the 
tensor product representations 
$(\pi_2\otimes \pi_1\otimes \pi_2\otimes \pi_1)\circ \Delta$
and 
$(\pi_2\otimes \pi_1\otimes \pi_2\otimes \pi_1)\circ \Delta^{\!\mathrm{op}}$
of $A_q(\mathrm{Sp}_4)$ defined by
\begin{align}
\Delta(t_{ij}) &= \sum_{1\le k,l,m \le 4}
\pi_2(t_{ik})\cdot \pi_1(t_{kl}) \cdot \pi_2(t_{lm}) \cdot \pi_1(t_{mj}),
\\
\Delta^{\!\mathrm{op}}(t_{ij}) &= \sum_{1\le k,l,m \le 4}
\pi_2(t_{mj}) \cdot \pi_1(t_{lm}) \cdot \pi_2(t_{kl}) \cdot \pi_1(t_{ik}),
\end{align}
where the symbol $\cdot$ is the abbreviation of  $\otimes$, and 
each component is given by
\begin{align}
\pi_1(t_{ij})= 
\begin{pmatrix}
\am & \nu_1\ok & 0& 0\\
-q\nu_1^{-1}\ok& \ap  & 0& 0\\
0 & 0 & \am & -\nu_1\ok \\
0 & 0 & q\nu_1^{-1}\ok & \ap
\end{pmatrix},\qquad
\pi_2(t_{ij})= 
\begin{pmatrix}
1  & 0 & 0 & 0\\
0 & \Am & \nu_2 \OK & 0\\
0 & -q^2\nu^{-1}_2 \OK & \Ap & 0\\
0 & 0 & 0  & 1\\
\end{pmatrix}.
\label{pi33}
\end{align}
By this we mean that the LHS is given by the element in the RHS at  
the $i$ th row and the $j$ th column from the top left. 
The parameters $\nu_1, \nu_2$ are free and do not influence the 
the subsequent argument, so we set $\nu_1 = \nu_2 = 1$ below.
The following was conjectured in \cite[eq.(78)]{KP}.

\begin{theorem}\label{th:bv}
\begin{align}\label{eig}
\mathscr{K}|\Xi_{r,k}\rangle = |\Xi_{r,k}\rangle
\qquad ((r,k)=(1,1), (1,2), (2,2)).
\end{align} 
\end{theorem}

By a direct calculation one can show
\begin{lemma}
\begin{alignat}{2}
1 \cdot \ok^2 \cdot \OK^2 \cdot 1 &= \Delta(t_{14}^2-q^{-3}t_{42}t_{13}),
\label{d1}\\
1 \cdot \ok \cdot \OK \cdot \ok &= \Delta(-t_{14})  =\Delta(q^{-4}t_{41}),
&\qquad
\OK \cdot \ok^2 \cdot \OK \cdot 1 &= \Delta(q^{-1}t_{23}t_{14}-t_{24}t_{13}),
\label{d2}\\
1 \cdot \ok \cdot \OK \cdot \ap  &= \Delta(-q^{-3} t_{42}),
&\qquad
1 \cdot \ok \cdot \OK \cdot \am  &= \Delta(t_{13}),
\label{d3}\\
\Ap \cdot \ok^2 \cdot \OK \cdot 1 &= \Delta(-q^{-5}t_{33}t_{41}-t_{34}t_{13}),
&\qquad
\Am \cdot \ok^2 \cdot \OK \cdot 1 &= \Delta(q t_{22}t_{14} + q^{-4}t_{21}t_{42}),
\label{d4}\\
1 \cdot \ok\, \ap  \cdot \OK \cdot 1 
&= \Delta(qt_{44}t_{13} + q^{-4}t_{43}t_{41}),
&\qquad
1 \cdot \ok\, \am \cdot \OK \cdot 1 
&= \Delta(-q^{-3}t_{42}t_{11} + q^{-4} t_{41}t_{12}),
\label{d5}\\
1\cdot \ok^2 \cdot \OK \Ap \cdot 1 
&= \Delta(-q^{-1} t_{44}t_{41} - q^{-2}t_{43}t_{42}),
&\qquad
1\cdot \ok^2 \cdot \OK \Am \cdot 1 
&= \Delta(-q^{-5}t_{41}t_{11} + q^{-2}t_{12}t_{13}).
\label{d6}
\end{alignat}
\end{lemma}

{\em Proof of Theorem \ref{th:bv}}.
In view of the definition (\ref{bv}) and the 
characterization (\ref{c11})--(\ref{c2}), it suffices to show
\begin{align}
((\Ap-\Am)\cdot \ok^2 \cdot \OK \cdot 1)
\mathscr{K}|\Xi_{2,2}\rangle &=0,
\label{h1}\\
(1 \cdot \ok\,(\ap-\am)  \cdot \OK \cdot 1)
\mathscr{K}|\Xi_{2,2}\rangle &=0,
\\
(1 \cdot \ok^2 \cdot \OK\,(\Ap-\Am)  \cdot 1)
\mathscr{K}|\Xi_{2,2}\rangle &=0,
\\
(1 \cdot \ok \cdot \OK \cdot (\ap-\am))
\mathscr{K}|\Xi_{2,2}\rangle &=0,
\\
((\Ap-\Am+(1+q^2)\OK)\cdot \ok^2 \cdot \OK \cdot 1)
\mathscr{K}|\Xi_{1,2}\rangle &=0,
\\
(1 \cdot \ok\,(\ap-\am)  \cdot \OK \cdot 1)
\mathscr{K}|\Xi_{1,2}\rangle &=0,
\\
(1\cdot \ok^2 \cdot \OK\,(\Ap-\Am+(1+q^2)\OK) \cdot 1)
\mathscr{K}|\Xi_{1,2}\rangle &=0,
\\
(1 \cdot \ok \cdot \OK \cdot (\ap-\am))
\mathscr{K}|\Xi_{1,2}\rangle &=0,
\\
((\Ap-\Am+(1+q^2)\OK)\cdot \ok^2 \cdot \OK \cdot 1)
\mathscr{K}|\Xi_{1,1}\rangle &=0,
\\
(1\cdot \ok\,(\ap-\am+(1+q)\ok) \cdot \OK \cdot 1)
\mathscr{K}|\Xi_{1,1}\rangle &=0,
\label{h2}\\
(1 \cdot \ok^2 \cdot \OK\,(\Ap-\Am+(1+q^2)\OK) \cdot 1)
\mathscr{K}|\Xi_{1,1}\rangle &=0,
\label{h3}\\
(1\cdot \ok \cdot \OK \cdot (\ap-\am + (1+q)\ok))
\mathscr{K}|\Xi_{1,1}\rangle &=0.
\end{align}
As an illustration, consider (\ref{h1}).
It is verified as 
\begin{align}
((\Ap-\Am)\cdot \ok^2 \cdot \OK \cdot 1)
\mathscr{K}|\Xi_{2,2}\rangle 
&\overset{(\ref{d4})}{=} \Delta(-q^{-5}t_{33}t_{41}-t_{34}t_{13}
-q t_{22}t_{14} - q^{-4}t_{21}t_{42})\mathscr{K}|\Xi_{2,2}\rangle\\
&\overset{(\ref{ir})}{=}\mathscr{K}\Delta^{\!\mathrm{op}}(
-q^{-5}t_{33}t_{41}-t_{34}t_{13}
-q t_{22}t_{14} - q^{-4}t_{21}t_{42})|\Xi_{2,2}\rangle = 0,
\end{align}
where the last equality is checked directly, although tedious,  by using 
(\ref{pi33}) and applying (\ref{c2}) to (\ref{bv}).  
The other relations can be shown similarly.
Namely one can always find a polynomial 
$T(\{t_{ij}\})$ which is a linear combination of those appearing  
in (\ref{d1})--(\ref{d6})  
such that the relation in question is expressed and shown as
\begin{align}
\Delta(T(\{t_{ij}\})) \mathscr{K}|\Xi_{r,k}\rangle
= \mathscr{K}\Delta^{\!\mathrm{op}}(T(\{t_{ij}\})) |\Xi_{r,k}\rangle = 0
\end{align}
by applying (\ref{c11})--(\ref{c2}) in the last step.

The only exception is (\ref{h2}) involving
$X = 1\cdot \ok^2 \cdot \OK \cdot 1$ which is {\em not} contained in the list 
(\ref{d1})--(\ref{d2}).
In fact, from (\ref{d5}), LHS of (\ref{h2}) is written as
\begin{align}\label{h4}
\left(
\Delta(
qt_{44}t_{13} + q^{-4}t_{43}t_{41} + 
q^{-3}t_{42}t_{11} - q^{-4} t_{41}t_{12}) + (1+q)X
\right)\mathscr{K}|\Xi_{1,1}\rangle.
\end{align}
To treat this, we rely on (\ref{h3}) which can be proved independently
as explained in the above.
It then tells that the third component of $|\Xi_{1,1}\rangle$
is proportional to $|\chi_1\rangle$.
Therefore from (\ref{c11}) we may claim that it also satisfies
\begin{align}
(1 \cdot \ok^2 \cdot  \OK(\Ap -1+ \OK) \cdot 1)\mathscr{K}|\Xi_{1,1}\rangle=0.
\end{align}
This leads to
\begin{align}
X\mathscr{K}|\Xi_{1,1}\rangle 
&= (1 \cdot \ok^2 \cdot  \OK(\Ap + \OK) \cdot 1)\mathscr{K}|\Xi_{1,1}\rangle
= \Delta(-q^{-1} t_{44}t_{41} - q^{-2}t_{43}t_{42}
+t_{14}^2-q^{-3}t_{42}t_{13})\mathscr{K}|\Xi_{1,1}\rangle
\end{align}
due to (\ref{d1}) and (\ref{d6}).
Substituting this into (\ref{h4}) and using (\ref{ir})  
one can check that it indeed vanishes.

\section*{Acknowledgments}
The authors thank Pascal Baseilhac for comments.
A.K. thanks Masato Okado and Akihito Yoneyama for 
collaboration in their previous works.
He is supported by Grants-in-Aid for Scientific Research 
No.~16H03922, 18H01141 and 18K03452 from JSPS.

\vspace{0.5cm}

\end{document}